\documentclass{aa}  

\usepackage{graphicx}
\usepackage{txfonts}
\newcommand{\vpeak}{V_{\rm peak}}
\newcommand{\vmax}{V_{\rm max}}
\newcommand{\MTNG}{MTNG}
\newcommand{\MTNGdmo}{MTNG-DMO}
\newcommand{\MTNGdmoMOD}{MTNG-DMO$\rm _{mod}$}
\newcommand{\Mr}{${\rm M}_r$}
\newcommand{\Mu}{${\rm M}_U$}
\newcommand{\hMsun}{ h^{-1}{\rm M_{ \odot}}}
\newcommand{\hMpc}{ h^{-1}{\rm Mpc}}

\newcommand{\ihMpcC}{ h^{3}{\rm Mpc}^{-3}}
\newcommand{\hkpc}{ h^{-1}{\rm kpc}}

\newcommand{\sig}{\sigma_{8}}
\newcommand{\OmM}{\Omega_\mathrm{M}}
\newcommand{\Omb}{\Omega_{\rm b}}

\newcommand{\ns}{{n_{\rm s}}}

\begin{document}

   \title{The effect of baryons on the positions and velocities of satellite galaxies in the MTNG simulation}

   %\subtitle{I. Overviewing the $\kappa$-mechanism}

   \author{Sergio Contreras
          \inst{1,2}\fnmsep\thanks{E-mail: scontreras1@us.es}
          \and
          Raul E. Angulo
          \inst{2,3}
           \and
           Sownak Bose
           \inst{4}
           \and
           Boryana Hadzhiyska
            \inst{5,6}
          \and
          Lars Hernquist
          \inst{7}
          \and
           Francisco Maion
          \inst{8,2}
          \and        
            Ruediger Pakmor
        \inst{9}
        \and        
        Volker Springel
        \inst{9}
          }

    \institute{
    Facultad de F\'isica. Universidad de Sevilla. Multidisciplinary Unit for Energy Science, Av. Reina Mercedes s/n 41012 Seville, Spain
    \and
    Donostia International Physics Center, Manuel Lardizabal Ibilbidea, 4, 20018 Donostia, Gipuzkoa, Spain
         \and
    IKERBASQUE, Basque Foundation for Science, 48013, Bilbao, Spain
    \and
    Institute for Computational Cosmology, Department of Physics, Durham University, South Road, Durham DH1 3LE, UK
    \and
    Kavli Institute for Cosmology Cambridge, Madingley Road, Cambridge, CB3 0HA, UK
    \and
    Institute of Astronomy, Madingley Road, Cambridge, CB3 0HA, United Kingdom
    \and
    Harvard-Smithsonian Center for Astrophysics, 60 Garden St, Cambridge, MA 02138, USA
    \and
    Columbia Astrophysics Laboratory, Columbia University, 550 West 120th Street, New York, NY 10027, United States of America
    \and
    Max Planck Institute for Astrophysics, Karl-Schwarzschild-Str. 1, D-85748 Garching, Germany
    }

   \date{Received March 14, 1592; accepted XXXX}

% \abstract{}{}{}{}{} 
% 5 {} token are mandatory
 
  \abstract
  % context heading (optional)
  % {} leave it empty if necessary  
    {Mock galaxy catalogues are often constructed from dark-matter-only simulations based on the galaxy–halo connection. Although modern mocks can reproduce galaxy clustering to some extent, the absence of baryons affects the spatial and kinematic distributions of galaxies in ways that remain insufficiently quantified. We compare the positions and velocities of satellite galaxies in the MTNG hydrodynamic simulation — a state-of-the-art cosmological hydrodynamic run — with those in its dark-matter-only counterpart, assessing how baryonic effects influence galaxy clustering and contrasting them with the impact of galaxy selection, i.e. the dependence of clustering on sample definition. Using merger trees from both runs, we track satellite subhaloes until they become centrals, allowing us to match systems even when their $z=0$ positions differ. We then compute positional and velocity offsets as functions of halo mass and distance from the halo centre, and use these to construct a subhalo catalogue from the dark-matter-only simulation that reproduces the galaxy distribution in the hydrodynamic run. Satellites in the hydrodynamic simulation lie 3–4\% closer to halo centres than in the dark-matter-only case, with an offset that is nearly constant with halo mass and increases toward smaller radii. Satellite velocities are also systematically higher in the dark-matter-only run, with differences that grow toward lower halo masses and radii. At scales of 0.1 $\hMpc$, these spatial and kinematic differences produce 10–20\% variations in clustering amplitude—corresponding to 1–3$\sigma$ assuming DESI-like errors—though the impact decreases at larger scales. We repeat the analysis in zoom-in simulations with varied physical models and find consistent trends. These baryonic effects are relevant for cosmological and lensing analyses and should be accounted for when building high-fidelity mocks. However, they remain smaller than the differences introduced by galaxy selection, which thus represents the dominant source of uncertainty when constructing mocks based on observable quantities.}

   \maketitle
%
%-------------------------------------------------------------------

\section{Introduction}
\label{sec:intro}
Galaxies form and evolve in dark matter haloes; therefore, we expect a strong correlation in the evolution of these two objects (\citealt{White:1978}). As haloes grow hierarchically, halo mergers become a fundamental way of increasing mass. While galaxies also merge, they do so on much longer timescales, so more massive haloes will contain several galaxies.

In a massive halo, the most massive and brightest galaxy is expected to be at the centre of the potential, with a velocity similar (but not identical) to that of the whole halo. This galaxy is commonly referred to as the central galaxy, whereas the remaining galaxies in the halo are known as satellite galaxies. When looking at the dark matter component of the halo, the position and velocity of satellites match those of subhaloes, which are the halo remnants of minor haloes that have fallen into the main halo. Using the Subhalo Abundance Matching formalism (SHAM, e.g. \citealt{Vale:2006, Conroy:2006}), subhaloes can be associated with galaxies based on their positions, velocities, and rotational-velocity history. This technique, often used in the creation of mock galaxy catalogues, is capable of reproducing several galaxy clustering statistics, such as the two-point correlation function (in real and redshift space; \citealt{ChavesMontero:2016, Ortega:2024}), galaxy–galaxy lensing (e.g. \citealt{Contreras:2023_LIL}), and even higher-order statistics (e.g. \citealt{Contreras:2024}), and has recently been used for the study of large-scale-structure effects such as assembly bias (e.g. \citealt{Contreras:2021_SHAMgab,Contreras:2023_SDSSlensing}), for characterising the clustering of the latest generation of galaxy surveys (e.g. \citealt{Rocher:2023, Ortega:2025}), and even for deriving cosmological constraints (e.g. \citealt{Contreras:2023_MTNG, Mahony:2025}). Mocks using the SHAM formalism are normally run based on dark-matter-only simulations. The low computational cost of these simulations allows running them on larger volumes and in large numbers, which is necessary when analysing the observations provided by surveys such as DESI or EUCLID (e.g. \citealt{ABACUSSUMMIT, EuclidFlagship01}).

While the SHAM approach is accurate and computationally efficient, it has some known limitations. The presence of baryons, for instance, can modify the halo density profiles as well as the positions and velocities of subhaloes inside haloes. Because empirical models (such as SHAMs or halo occupation models) are usually calibrated on dark-matter-only simulations, and the positions and velocities of galaxies are set to be equal to those of subhaloes, dark matter particles, or to follow a standard NFW profile, the absence of baryonic effects will impact the clustering signal of these galaxies.

Modern abundance-matching models use additional criteria to determine which satellite subhaloes should be populated by a specific type of galaxy (\citealt{Contreras:2021_SHAMe}). These extensions are especially useful for accounting for selection effects, which refer to the fact that galaxy samples are built using different observational criteria. For instance, blue satellite galaxies are situated in the outer regions of haloes (\citealt{Orsi:2018}), whereas redder galaxies are more prevalent in the inner regions. By adding this extra flexibility, models can populate subhaloes that better resemble the distribution of galaxies selected by a specific property (such as a cut in stellar mass or star formation rate). These mechanisms can also partially mimic the effects of baryons in the satellite distribution. While they do not shift the position of the dark matter subhalo as baryons do, they will populate subhaloes that are closer to the inner or outer part of the halo, producing a similar effect. However, to our knowledge, no comprehensive study has jointly quantified the impact of baryons on the positions and velocities of satellite galaxies and compared it directly to the impact of selection effects.

In this paper, we make use of a state-of-the-art cosmological hydrodynamic simulation, the MillenniumTNG simulation (\MTNG, \citealt{MTNG_01,MTNG_02}), to measure the impact of baryons on the positions and velocities of satellite subhaloes. By comparing the galaxy population of the hydrodynamic simulation (which models baryons and dark matter simultaneously) to that of a dark-matter-only simulation, run with the same initial conditions and cosmology as \MTNG, we can determine, satellite by satellite, the shift in position and velocity caused by baryons. The large volume and high resolution of \MTNG\ provide a unique opportunity to perform this comparison over a broad halo-mass range with a large number of well-resolved satellites, thereby ensuring accurate measurements of halo profiles.

Matching central objects between a dark-matter-only and a full hydrodynamic simulation (both run with the same initial conditions) is quite simple, since we expect the total mass and position of the haloes to be similar. Unfortunately, this is not the case for satellite galaxies. Once a central becomes a satellite, due to several processes, including baryonic effects, the pair of satellites will no longer share either a similar position or mass. To match satellite galaxies robustly, we track each individual object to a previous snapshot until it becomes a central and has a mass close to its peak halo mass during its evolutionary history. This process allows us to successfully match more than 97.5\% of the satellite galaxies above the minimum stellar-mass limit we set for this study. We find that satellite galaxies in the hydrodynamic simulation are $\sim$3\% closer to the centre of their halo than subhaloes in the dark-matter-only simulation, and also move up to $\sim$3\% slower (with a small dependence on their halo mass). We show that this translates into a difference of up to 10\% in galaxy clustering at scales of $\sim 0.1 \hMpc$. While definitely significant, these differences are much lower than those we would find when selecting galaxies using a threshold based on different galaxy properties, such as stellar mass, luminosity in a given band, or star formation rate. We therefore conclude that, while baryonic effects significantly affect galaxy clustering and should be taken into account in precision cosmology, their impact is superseded by selection effects, which should be the priority when improving the modelling of galaxies in dark-matter-only simulations. 

We use a suite of zoom-in hydrodynamic simulations, each implementing modified baryonic physics, to assess how such changes affect satellite properties.  Although the limited number of massive haloes prevents robust statistical constraints, our results indicate that, even under extreme models, the main trends reported in this paper are unlikely to vary by more than a factor of two.

The outline of this paper is as follows. Section~\ref{sec:numeric} presents the \MTNG\ simulation, as well as its dark-matter-only counterpart, and describes the galaxy and subhalo samples used, along with the galaxy-matching algorithm. The impact of the baryonic effects on the positions and velocities of subhaloes is shown in Sections~\ref{sec:pos} and~\ref{sec:vel}, respectively. Section~\ref{sec:clustering} shows how these differences in positions and velocities propagate into the galaxy correlation function. We finalise by presenting our conclusions in Section~\ref{sec:conclusions}. Appendix~\ref{sec:Mimicking} describes the algorithm used to modify the positions and velocities of subhaloes in the dark-matter-only simulation so as to emulate the impact of baryons, while Appendix~\ref{sec:zoom} examines how our main results change under alternative implementations of baryonic physics.

Unless otherwise stated, the standard units in this paper are $\hMsun$ for masses and $\hMpc$ for distances. All logarithm values are in base 10.

\section{Numerical simulation and galaxy samples}
\label{sec:numeric}

In this section, we present the cosmological hydrodynamic simulation used in this work, the MillenniumTNG (\MTNG), as well as its dark-matter-only counterpart (Section~\ref{sec:MTNG}). Section~\ref{sec:matched} describes the algorithm employed to match objects between the two simulations. Finally, in Section~\ref{sec:samples}, we outline the criteria used to select galaxies in each sample for comparison.
\subsection{The MTNG simulations}
\label{sec:MTNG}

To quantify the impact of baryons on satellite galaxies, we require a hydrodynamic simulation with a large volume to ensure accurate statistics, as well as high resolution to resolve several satellites per halo. One of the few simulations available today with these characteristics is the largest hydrodynamic run of the MillenniumTNG suite of simulations \citep{MTNG_01,MTNG_02,MTNG_03,MTNG_04,MTNG_05,MTNG_06,MTNG_07,MTNG_08,Contreras:2023_MTNG}. This suite comprises tens of dark-matter-only and hydrodynamic simulations, along with light-cone galaxy catalogues generated using semi-analytical models. The project extends the two well-known Millennium \citep{Springel:2005} and IllustrisTNG \citep{TNGa,TNGb,TNGc,TNGd,TNGe,Nelson:2019} projects in a direction that enables accurate studies of the galaxy–halo connection and of the impact of baryonic physics on clustering, particularly at much larger cosmological volumes than previously possible.

The largest hydrodynamic simulation of this suite (which we refer to simply as the \MTNG\ simulation) has a periodic volume of $(500\ \hMpc)^3$. It contains $4320^3$ dark matter particles and an equal number of gas cells, implying an average gas-cell mass of $2.00\times10^7\,\hMsun$ and a mass resolution of $1.12\times10^8\,\hMsun$ for the dark matter particles. The simulation adopts a \cite{Planck:2015} cosmology\footnote{$\OmM = 0.3089$, $\Omb = 0.0486$, $\sig = 0.8159$, $\ns = 0.9667$, and $h = 0.6774$}, identical to that used in the IllustrisTNG project.

The initial conditions for \MTNG\ were generated by fixing the amplitudes of the initial power modes to their expected {\it rms} values \citep{Angulo:2016}, which substantially reduces the effect of cosmic variance, at least for second-order statistics. The simulation was run using the moving-mesh code \texttt{AREPO} \citep{AREPO}, which includes radiative cooling, star formation in the gas, the growth of supermassive black holes, and the associated feedback from both supernovae and active galactic nuclei. Galaxies form as agglomerations of star particles whose properties can be directly measured from the simulation. The run comprises 265 snapshots,  ranging from $z=63$ (when the initial conditions were generated) to $z=0$. Due to space limitations, only the group catalogues were stored, except for a limited number of key snapshots that also included particle data. This number of group and subhalo outputs is several times greater than in most previous large-scale simulations, providing a unique opportunity to follow the detailed evolution of satellite galaxies within haloes.

To study the impact of baryons on satellite subhaloes, we need to match the galaxies from \MTNG\ to a simulation with similar characteristics but composed only of dark matter. For this purpose, we use another simulation from the MillenniumTNG suite with the same number of dark matter particles, the same cosmology, and the same initial conditions and volume. Both simulations were run with the moving-mesh code \texttt{AREPO} and have identical snapshot outputs at the same redshifts. We refer to this simulation as \MTNGdmo. Although we use a simulation as similar as possible to the original \MTNG\ run to minimise noise in our comparison, we also tested matching our galaxy sample to a lower-resolution simulation (with $2^3$ fewer particles) that shares the same cosmology, initial conditions, and number of snapshots as \MTNG, but was run with the \texttt{Gadget-4} code \citep{Gadget-4}. We found only minor differences in our main results.

\subsection{Matched catalogues}
\label{sec:matched}

Because \MTNG\ and \MTNGdmo\ share identical initial conditions (ICs), most haloes in one simulation can identify a ``partner'' object in the other with similar mass, position, and formation history, which originated from the same overdensity in the ICs. To match the host haloes (not subhaloes or galaxies) between the two simulations, we employ a kd-tree algorithm from \texttt{SciPy} \citep{scipy} to search for neighbouring objects in a four-dimensional space defined by position and logarithmic halo mass. We set the weight of being off by one dex in halo mass to be equal to being off by 1 $\hMpc$. We test other configurations, only finding minor differences and a noticeable change in our results. To ensure robust matched catalogues, the procedure is performed bi-directionally: we first search for haloes in \MTNGdmo\ that best match those in \MTNG, and then repeat the search in the opposite direction.

It is not possible, however, to apply this same matching algorithm directly to satellite galaxies or subhaloes. Due to both numerical and physical effects (such as baryonic processes), satellites gradually deviate from their counterparts in the other simulation, and dynamical friction cause their masses to decrease at different rates. To overcome this, we use the merger trees of both simulations to trace the main progenitor branch of all satellite galaxies in \MTNGdmo\ back to the snapshot where they were still central objects, with masses of approximately $80\%$ of their peak subhalo mass (i.e. the maximum mass reached during their evolution). We then repeat the matching procedure described above for the host haloes at this specific snapshot. This process is applied to all satellite galaxies residing in haloes more massive than $10^{12}\,\hMsun$ and with peak maximum circular velocities above $100\,\mathrm{km\,s^{-1}}$ (corresponding to the lowest cut in our selection samples; see Section~\ref{sec:samples} for details). We adopt $80\%$ of the peak subhalo mass because, at this stage, the objects are still massive enough for reliable identification by our algorithm, yet not so close to the haloes into which they will eventually merge that the matching becomes ambiguous. We tested alternative thresholds and found only minor differences in our results. This technique successfully matches more than $97.5\%$ of all objects in our sample.

\subsection{Target galaxy samples}
\label{sec:samples}

\begin{figure}
\includegraphics[width=0.45\textwidth]{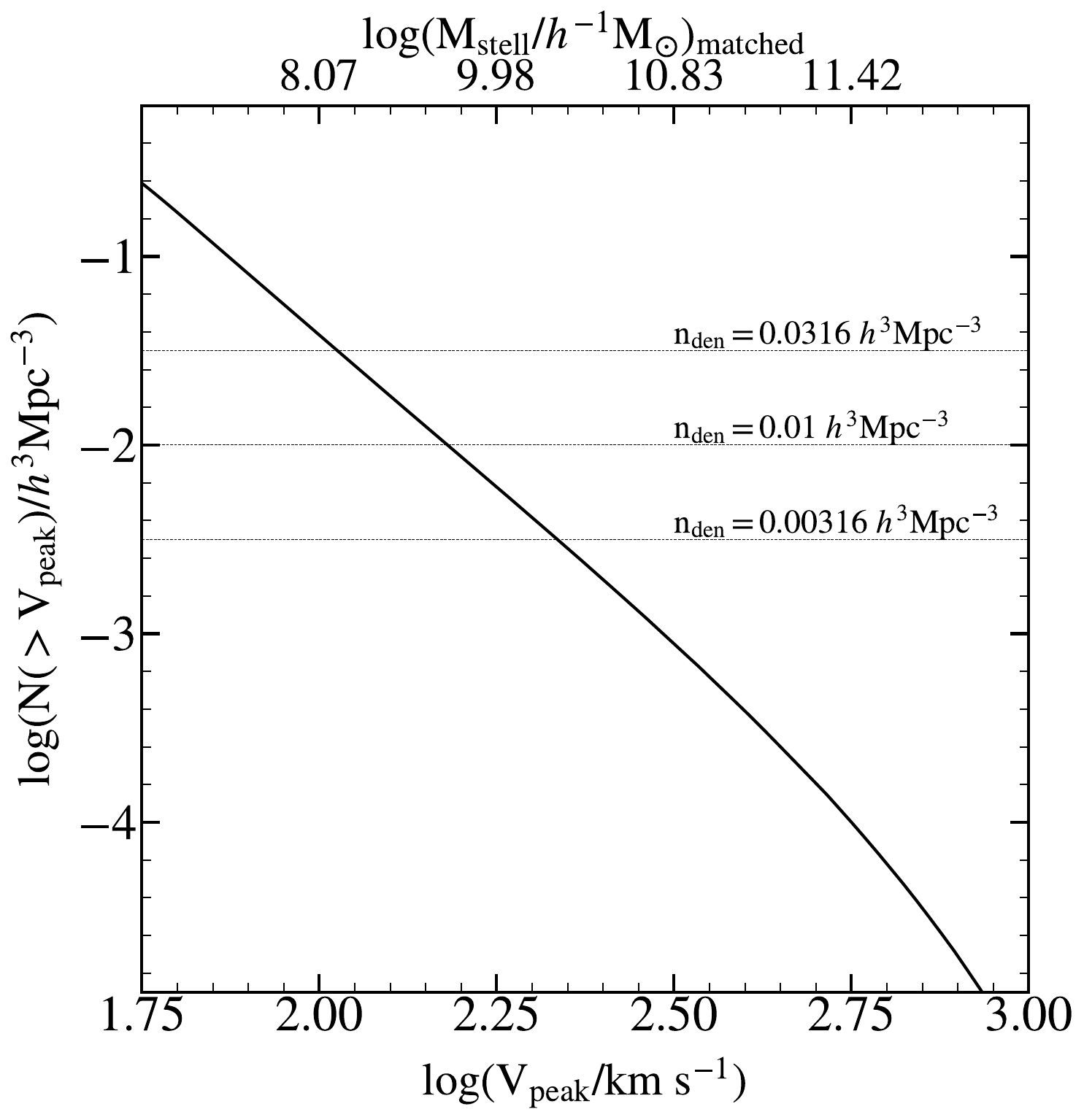}
\caption{Cumulative distribution of $\vpeak$ for subhaloes in the \MTNGdmo\ simulation. We define $\vpeak$ as the maximum circular velocity ($\vmax$) reached by a subhalo during its lifetime. The upper axis shows the corresponding stellar mass of these haloes after matching them to their counterparts in the hydrodynamic \MTNG\ simulation.}
\label{Fig:vpeak_funct}
\end{figure}

To study the effects of baryons on satellite subhaloes, we selected subhaloes directly from the dark-matter-only simulation and matched them to their counterparts in the hydrodynamic simulation. We chose to select subhaloes from the dark-matter-only run for simplicity, as it contains fewer objects (probably caused by the lower mass of the stellar mass particle, and that they are usually found in the inner cores of the subhaloes), and because this facilitates the characterisation of baryonic effects in the type of simulation most commonly used to build mock galaxy catalogues. The subhaloes were selected according to the peak value of their maximum circular velocity, $\vmax \equiv \sqrt{GM(<r)/r}$, attained during their evolution ($\vpeak$). As shown by \cite{ChavesMontero:2016}, subhaloes selected by $\vpeak$ exhibit clustering properties similar to those of galaxies selected by stellar mass, making it a standard choice for creating mocks in SHAM-like models. As we will see in the following sections, baryonic effects are stronger in the inner regions of haloes. Selecting galaxies by stellar mass or $\vpeak$ also preferentially targets objects located in these inner regions (in contrast to star-forming samples, which are mainly found in the outer regions of haloes; \citealt{Orsi:2018}). Consequently, this selection is expected to be sensitive to baryonic effects, making $\vpeak$ an appropriate variable for defining our samples.

The subhalo samples were constructed at $z=0$ for three different number densities: $0.0316$, $0.01$, and $0.00316~\ihMpcC$. A graphical representation of these samples is shown in Fig.~\ref{Fig:vpeak_funct}, where each sample corresponds to the subhaloes located to the right of the intersection between the horizontal lines and the cumulative $\vpeak$ function. To facilitate comparison with a galaxy sample selected by stellar mass, the upper panel of the figure indicates the equivalent stellar mass for subhaloes with ${\rm log}(\vpeak)=2.00$, $2.25$, $2.50$, and $2.75$, assuming the stellar mass function of \MTNG\ and a one-to-one mapping between stellar mass and $\vpeak$ in the dark-matter-only simulation (i.e. the most massive galaxy in the hydrodynamic simulation is assigned the largest $\vpeak$ value in the dark-matter-only simulation, the second most massive galaxy is assigned the second largest value, and so on, as in a basic SHAM approach).

For the final part of this paper, we also select galaxies directly from the hydrodynamic \MTNG\ simulation based on their stellar mass (defined as the total mass of stellar particles within twice the stellar half-mass radius), luminosity in the $r$ and $U$ bands (computed as the summed luminosities of all stellar particles in the subhalo), and star formation rate (SFR). These samples are defined to have the same number densities as the subhalo samples selected by $\vpeak$.

\section{Impact of baryons on the position of galaxies}
\label{sec:pos}

\begin{figure}
\includegraphics[width=0.45\textwidth]{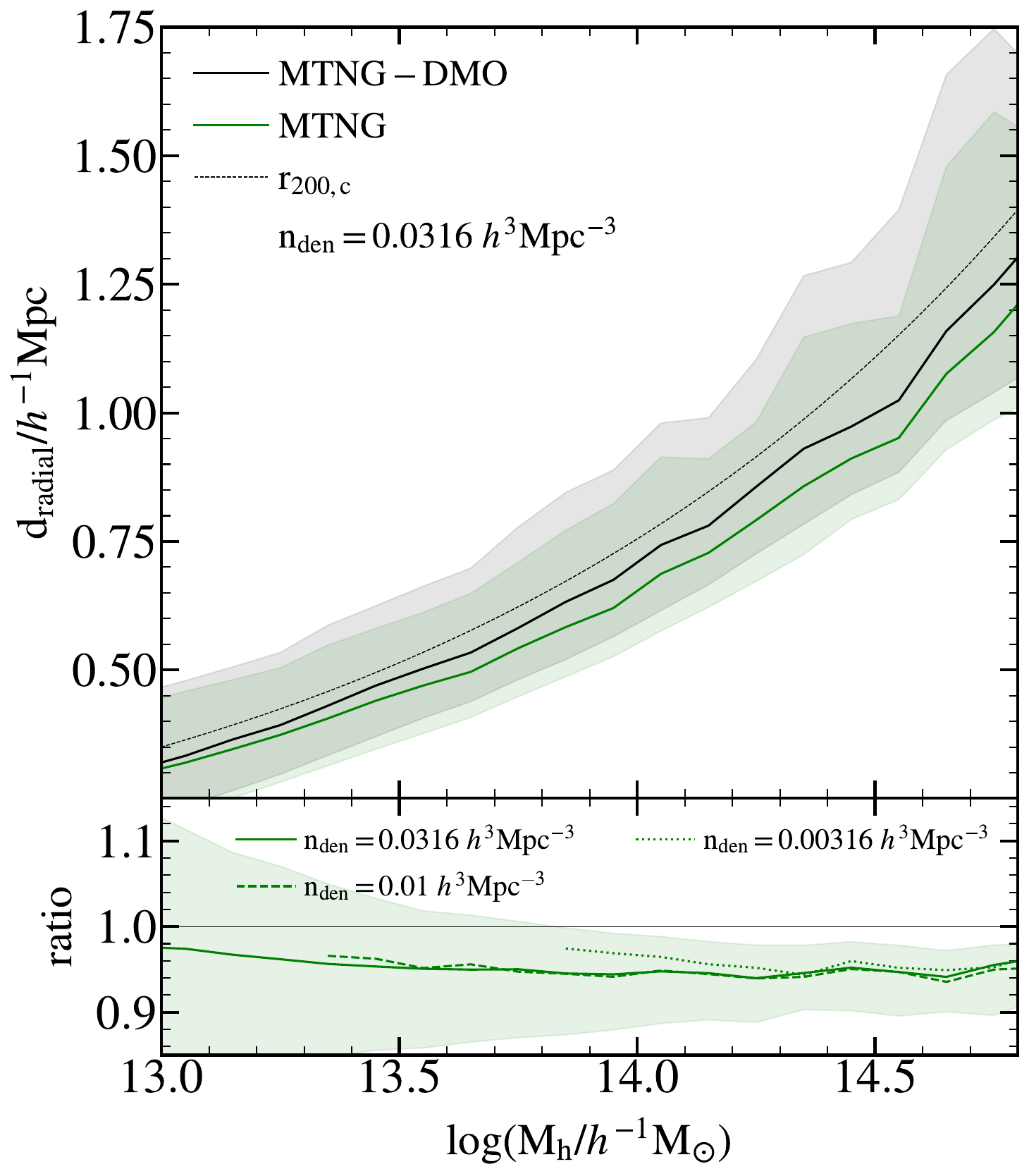}
\caption{(Top) Median (solid lines) and 16th–84th percentile region (shaded area) of the mean radial distance between satellite galaxies and the halo centre as a function of halo mass. Results from the \MTNGdmo\ simulation are shown in black, and those from the hydrodynamic \MTNG\ run in green. Subhaloes were selected according to their $\vpeak$ for a number density of ${\rm n_{den}=0.0316}~\ihMpcC$ in \MTNGdmo, and the corresponding matched galaxies in \MTNG\ were identified following the procedure described in Section~\ref{sec:matched}. Distances are shown as a function of the \MTNGdmo\ halo masses. The black dashed line represents the value of one virial radius $r_{200c}$. (Bottom) Median (solid lines) and 16th–84th percentile region (shaded area) of the ratio of mean distances between matched haloes in the \MTNG\ and \MTNGdmo\ simulations. The green solid line and shaded area correspond to the galaxy sample with number density $0.0316~\ihMpcC$ (same as in the top panel), while the dashed and dotted lines represent galaxy samples with number densities of $0.01$ and $0.00316~\ihMpcC$. Only the subhaloes successfully matched between the two simulations are shown here.}
\label{Fig:mean_dist}
\end{figure}

\begin{figure}
\includegraphics[width=0.40\textwidth]{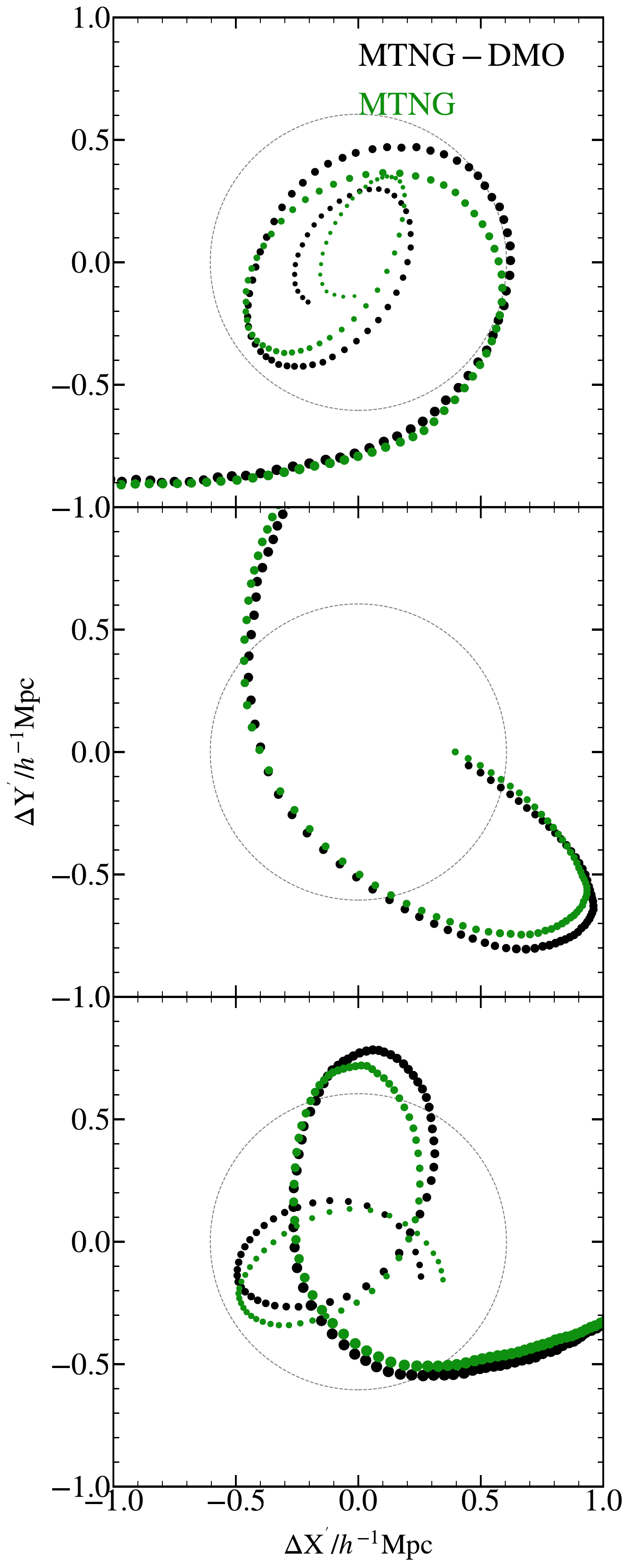}
\caption{Illustration of three satellite galaxies infalling into a halo of $\sim10^{13.5},\hMsun$. Black circles show the satellite positions in the \MTNGdmo\ simulation, while green circles indicate their matched counterparts in the hydrodynamic \MTNG\ run. The symbol sizes scale with the mass of the satellites. The dashed grey line represents one virial radius. }
\label{Fig:orbits}
\end{figure}

\begin{figure*}
\includegraphics[width=0.95\textwidth]{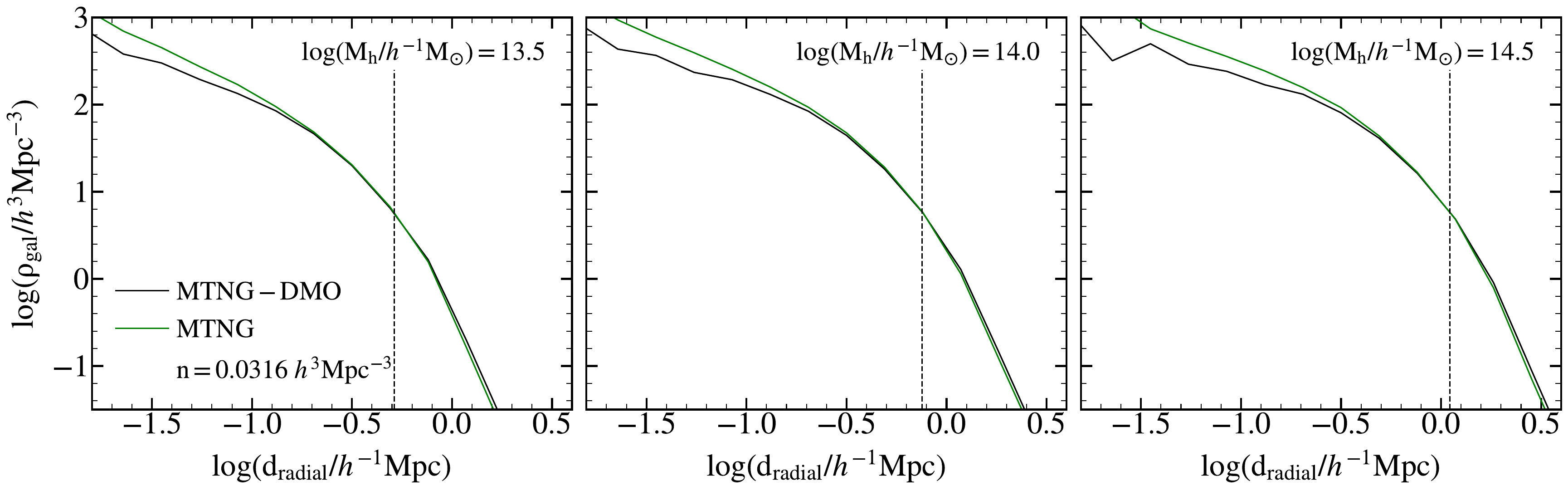}
\caption{Density of satellite galaxies as a function of distance from the halo centre for the $\rm n_{den}=0.0316~\ihMpcC$ sample. The left, middle, and right panels show density profiles for halo masses of $10^{13.5}$, $10^{14}$, and $10^{14.5}~\hMsun$, respectively. Black lines represent results from the \MTNGdmo\ simulation, while green lines correspond to the hydrodynamic \MTNG\ simulation. The vertical dashed lines represent the value of one virial radius $r_{200c}$.}
\label{Fig:prof_dist}
\end{figure*}

We now examine the positional differences between the $\vpeak$-selected satellite subhaloes of \MTNGdmo\ and their matched galaxies in the hydrodynamic \MTNG\ simulation. The upper panel of Fig.~\ref{Fig:mean_dist} shows the mean distance of satellites from their halo centres as a function of halo mass for the $\rm n_{den}=0.0316~\ihMpcC$ sample. The solid line represents the median of the distribution, and the shaded region indicates the 16–84\% range. Satellite galaxies in the hydrodynamic simulation are found to be 3–4\% closer to their halo centres than their counterparts in the dark-matter-only simulation. A reduction in satellite distances is consistent with a scenario in which satellites lose kinetic energy and reduce their orbital angular momentum as they fall inward due to dynamical friction \citep{Ogiya:2016}. Ram-pressure forces also contribute for galaxies that retain some of their gas, since gas experiences a stronger drag than dark matter. Through gravitational coupling within the satellite, this force is effectively transmitted to the system as a whole. In addition, satellites in the hydrodynamic simulation experience gaseous drag (i.e., the general retarding force on an object moving through gas), which differs from ordinary dynamical friction against the dark-matter background. For slightly supersonic motion—a regime commonly encountered by satellites—the gas drag can be even stronger \citep{Ostriker:1999}. Two other effects that can cause differences between a dark-matter-only run and one including baryons are tidal shocking —in particular, disk shocking, which can lead to satellite depletion in the inner regions of halos \citep{DOnghia:2010}— and modifications to the host’s density profile and the associated strengthening of tidal forces due to the presence of a central galaxy.The combination of these mechanisms leads to a progressive loss of orbital angular momentum, which becomes most noticeable after each pericentric passage \citep{Miller:2020}. Consistently, when examining individual orbits, we find that the discrepancy between the simulations emerges after the first pericentric passage. An example of three such satellite orbits is shown in Fig.~\ref{Fig:orbits}.

Figure~\ref{Fig:mean_dist} also shows no significant dependence of this offset on halo mass. The bottom panel of the figure, which presents the ratio between the mean satellite distances in the \MTNG\ and \MTNGdmo\ simulations, includes the lower-density samples and reveals no notable change in trend across different number densities. To minimise noise, we restrict the analysis in this panel to haloes that host, on average, at least five satellites per halo (i.e. $\rm \langle N_{sat}(M_{h,\min}) \rangle > 5$). Lower-mass haloes display similar behaviour (not shown here).

We next explore how the baryonic effects depend on distance from the halo centre. Figure~\ref{Fig:prof_dist} shows the mean satellite density profiles for haloes with $\log(M_h/\hMsun) \simeq 13.5$, 14, and 14.5 in the \MTNG\ and \MTNGdmo\ simulations. The halo mass bins have a width of 0.1 dex. As in the previous figure, the abundance of galaxies in the hydrodynamic simulation is higher at smaller radii than in the dark-matter-only run. The difference between simulations becomes most apparent on small scales, where the abundance of satellites in \MTNG\ exceeds that in \MTNGdmo\ by up to $\sim10\%$ at radii of $\sim30\,h^{-1}{\rm kpc}$, largely independent of halo mass. Only the highest-density sample is shown here, as the other samples exhibit similar trends. These results are consistent with those found by \cite{Vogelsberger:2014a} and \cite{Vogelsberger:2014b} in the original Illustris hydrodynamic simulation.

The enhanced abundance of satellites in the inner regions of host haloes may result from the increased central density —particularly of gas— on these scales. As satellites pass through pericentre, they lose orbital energy and slow down, bringing them closer to the halo centre and increasing the time they spend in this region, thereby affecting their radial distribution. We will examine this interpretation further in the next section.

\section{Impact of baryons on the velocities of galaxies}
\label{sec:vel}

\begin{figure}
\includegraphics[width=0.45\textwidth]{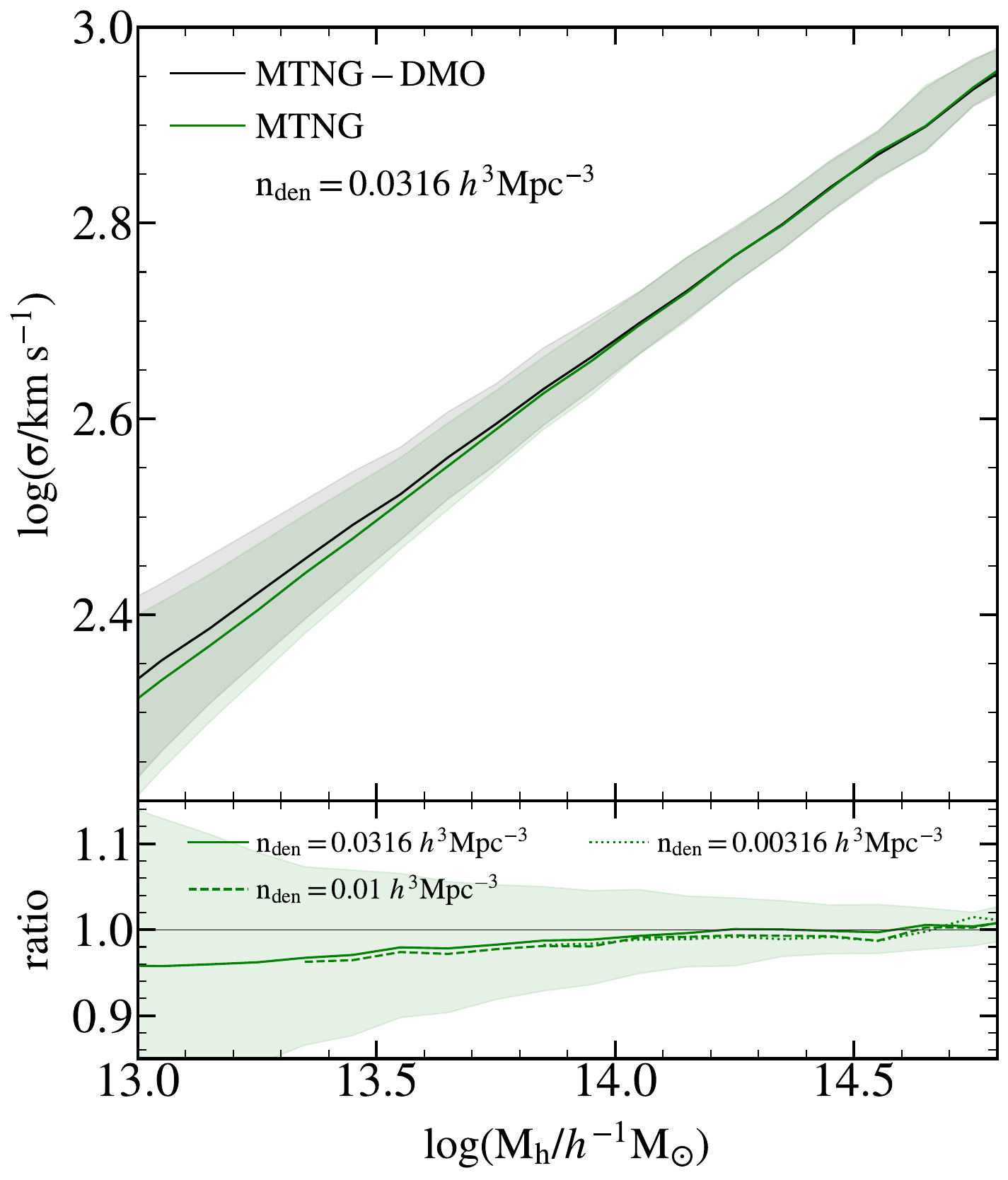}
\caption{(Top) Median (solid lines) and 16th–84th percentile region (shaded area) of the satellite velocity dispersion as a function of halo mass for a number density of $\rm n_{den}=0.0316~\ihMpcC$. Results from the \MTNGdmo\ and \MTNG\ simulations are shown in black and green, respectively. (Bottom) Median (solid lines) and 16th–84th percentile region (shaded area) of the ratio of velocity dispersions between matched haloes in the \MTNG\ and \MTNGdmo\ simulations, for the three number densities used in this work (as labelled).}
\label{Fig:mean_vel}
\end{figure}

\begin{figure*}
\includegraphics[width=0.95\textwidth]{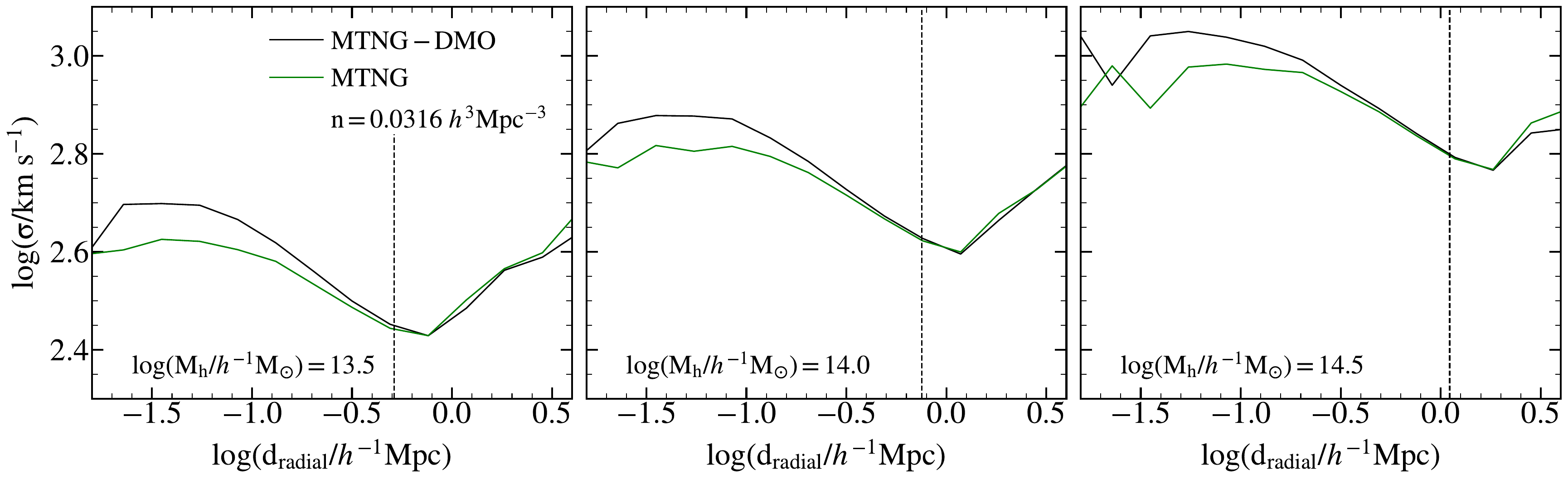}
\caption{Mean velocity dispersion of satellite galaxies as a function of distance from the halo centre for the $\rm n_{den}=0.0316~\ihMpcC$ sample. The left, middle, and right panels show velocity profiles for halo masses of $10^{13.5}$, $10^{14}$, and $10^{14.5}~\hMsun$, respectively. Black lines represent results from the \MTNGdmo\ simulation, while green lines correspond to the hydrodynamic \MTNG\ run. The vertical dashed lines represent the value of one virial radius $r_{200c}$.}
\label{Fig:prof_vel}
\end{figure*}

We now focus on the velocities of satellite galaxies within haloes. To quantify the differences between simulations, we compute the relative velocity of each satellite with respect to its host halo, $v'_{\mathrm{sat}} = v_{\mathrm{sat}} - v_{\mathrm{h}}$. We then define the satellite velocity dispersion as the three-dimensional velocity dispersion of the satellites within a halo.

Figure~\ref{Fig:mean_vel} shows the median and 16–84\% range of the mean satellite velocity dispersion in \MTNG\ and \MTNGdmo\ as a function of halo mass. In contrast to the behaviour observed for distances to the halo centre, the mean velocity dispersion is nearly identical for the most massive haloes. For less massive haloes, we find small differences of a few per cent, with satellites in the dark-matter-only simulation moving slightly faster than in the hydrodynamic run. The agreement across the three number densities (bottom panel) is even better than for positions, showing no systematic trend across samples.

Although the overall velocity dispersion is very similar between the two simulations, a clear difference appears when considering the dependence on distance from the halo centre, as shown in Fig.~\ref{Fig:prof_vel}. In the inner regions of haloes, satellite velocities are up to $\sim10\%$ lower in the hydrodynamic simulation. This difference is primarily caused by dynamical friction that slow down satellites as they pass through pericentre.

The decrease in velocity within the central regions is present at all halo masses, which may seem at first inconsistent with Fig.~\ref{Fig:mean_vel}, where we find comparable mean velocity dispersions for the most massive haloes. This apparent discrepancy arises because satellites in the hydrodynamic simulation move more slowly in the inner regions but remain there for longer periods, as seen in Fig.~\ref{Fig:prof_dist}. Although their instantaneous velocities are lower, these satellites still move faster than those in the outer halo regions. The two effects compensate, keeping the mean satellite velocity nearly constant for massive haloes and producing a small increase in velocity in the dark-matter-only simulation toward lower masses. This interpretation is consistent with feedback processes being more effective at lower halo masses—particularly supernova feedback, while AGN feedback remains active \citep{Akino:2022}—and thus with stronger baryonic effects in the inner parts of the mass density profile. After the first pericentric passage, satellites retain slightly lower velocities, as shown in the $0.1$–$1.0~\hMpc$ range of Fig.~\ref{Fig:prof_vel}. In this regime, the population includes recently infalling haloes, so the velocity suppression is less pronounced. Beyond $\sim1~\hMpc$, the velocity distributions of the two simulations converge, as these objects are about to begin infall or are nearby systems identified as satellites by the halo finder, which we do not consider further in this analysis.

We conclude from these results that baryons directly affect satellite velocities, with the strongest influence occurring near pericentric passage. This effect also modifies the spatial distribution of satellites within haloes, drawing them closer to the centre and increasing the time they spend in the inner regions. Because satellites move faster in the inner parts of haloes, the median (and mean) velocity dispersion of the system does not decrease substantially, even though individual satellites experience lower orbital speeds along their trajectories. Therefore, this statistic should be interpreted with care when assessing the impact of baryons on satellite dynamics.

\section{Baryonic and selection effects on galaxy clustering}
\label{sec:clustering}

\begin{figure*}
\includegraphics[width=0.95\textwidth]{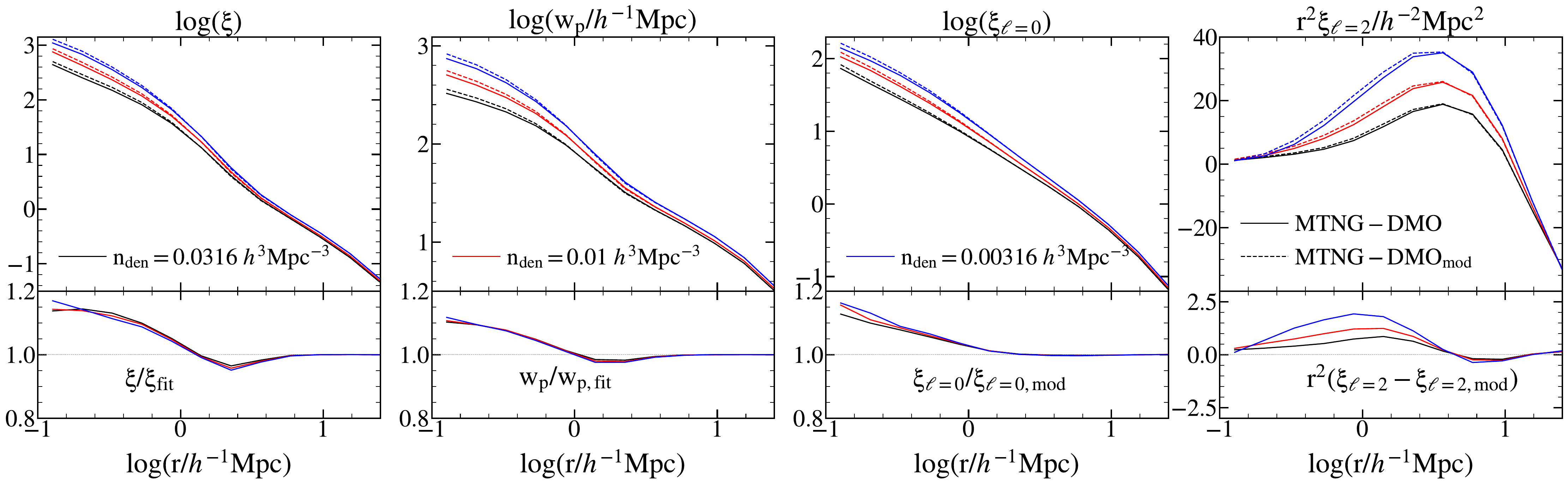}
\caption{(Top) From left to right: the real-space two-point correlation function, the projected correlation function, the monopole, and the quadrupole of the correlation function for samples with number densities of 0.0316 (black lines), 0.01 (red lines), and 0.00316~$\ihMpcC$ (blue lines). Solid lines represent the standard \MTNGdmo\ samples, while dashed lines correspond to the clustering from the modified \MTNGdmo\ simulation, where satellite positions and velocities were adjusted to mimic those of the hydrodynamic \MTNG\ run (\MTNGdmoMOD). (Bottom) Ratio of the clustering between \MTNGdmoMOD\ and \MTNGdmo\ for the two-point, projected, and monopole correlation functions. For the quadrupole (rightmost panel), we show the difference rather than the ratio to avoid discontinuities when the signal crosses zero.}
\label{Fig:clustering}
\end{figure*}

\begin{figure*}
\includegraphics[width=0.95\textwidth]{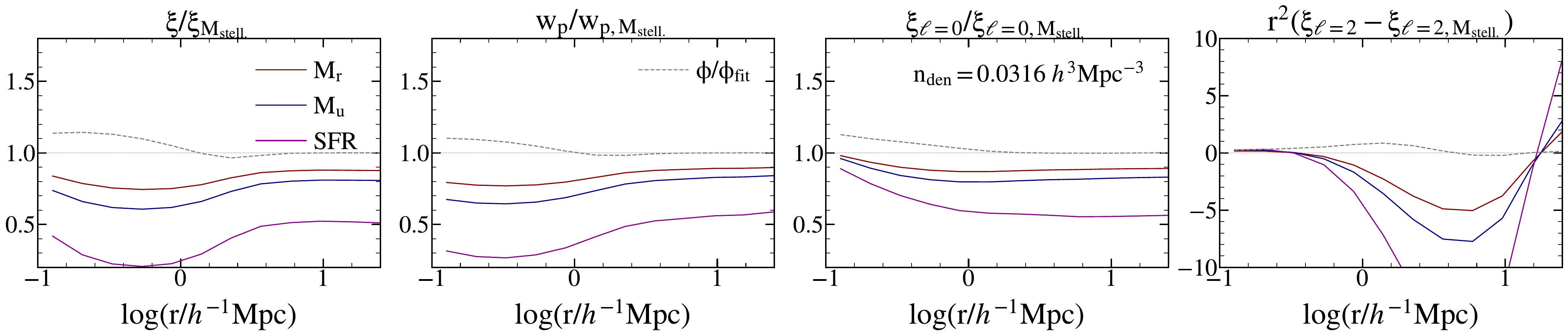}
\caption{Ratio and difference in galaxy clustering between galaxies selected by $\rm M_r$ luminosity (dark red lines), $\rm M_u$ luminosity (dark blue lines), and star formation rate (SFR, magenta lines), and those selected by stellar mass in the \MTNG\ simulation. Each galaxy sample has a number density of 0.0316~$\ihMpcC$. For comparison, dashed grey lines show the ratio and difference in clustering between $\vpeak$-selected subhaloes in the \MTNGdmoMOD\ and \MTNGdmo\ simulations (equivalent to the black solid line in the bottom panel of Fig.~\ref{Fig:clustering}).}
\label{Fig:SelEff}
\end{figure*}

In this section, we examine the impact of baryonic effects on several two-point statistics. Section~\ref{sec:clustering_eff} describes how we modified the dark-matter-only simulation to reproduce the baryonic effects observed in the hydrodynamic run and presents their impact on subhalo clustering. A more technical description of the baryon-mimicking procedure is provided in Appendix~\ref{sec:Mimicking}. In Section~\ref{sec:sel_effect}, we compare the magnitude of baryonic effects with those introduced by galaxy-selection criteria.

\subsection{The impact of baryons on galaxy clustering}
\label{sec:clustering_eff}

In the previous sections, we explored the direct impact of baryons on satellite positions and velocities, finding variations of up to 10\% on small scales. We now focus on how these differences affect galaxy clustering in real and redshift space. So far, our analysis has been based on subhaloes in \MTNGdmo\ above $10^{12}\,\hMsun$ that could be matched to the hydrodynamic \MTNG. Although the completeness of this matching exceeds 97.5\%, computing clustering from an incomplete sample can introduce systematic biases. To account for this, we parameterise the positional and velocity differences found in Sections~\ref{sec:pos} and~\ref{sec:vel}, and apply corresponding corrections to ensure that all subhaloes in the dark-matter-only simulation follow the same trends as in the hydrodynamic run. Below, we briefly describe these corrections; further details are given in Appendix~\ref{sec:Mimicking}.

\subsubsection*{Change in the positions of satellites}

We shift the positions of satellite subhaloes in \MTNGdmo\ to reproduce the radial distribution of galaxies in \MTNG. The cumulative satellite density profile is computed for both simulations as a function of distance from the halo centre, normalised by the virial radius. This statistic is found to be nearly independent of halo mass. We then derive the mean radial displacement as a function of cumulative satellite abundance and apply this relation to all satellite subhaloes in \MTNGdmo, effectively reducing their distances from the halo centres.

\subsubsection*{Change in the velocities of satellites}

To correct for velocity differences, we measure the ratio of satellite velocity dispersions between the simulations as a function of halo mass (analogous to Fig.~\ref{Fig:prof_vel}). This ratio is well approximated by an error-function-like expression, equal to 1 on large scales and approaching 0.9 at small scales, with a smooth transition between the two. The small-scale value shows a weak dependence on halo mass, which we include in the model. We then apply this function to all subhaloes in the dark-matter-only simulation, thereby reducing their velocities to match the distribution in the hydrodynamic run.

It is essential to apply this velocity correction \emph{after} the positional shift; otherwise, the median velocity of the sample cannot be correctly recovered, as discussed in Section~\ref{sec:vel}.

\subsubsection*{Change in the velocities of centrals}

While the positions of central subhaloes and central galaxies coincide with the halo centre (i.e. the potential minimum), their velocities differ from that of the halo itself—typically defined as the mass-weighted mean velocity of all particles or cells in the group. This offset, known as the velocity bias, correlates with halo velocity dispersion and can significantly influence galaxy clustering \citep{Guo:2015}. By construction, our use of central subhaloes from the dark-matter-only simulation implicitly includes velocity bias; however, its amplitude may differ from that in the hydrodynamic simulation. To create an equivalent galaxy sample and minimise systematic differences, we correct for this bias.

We first match all haloes above $10^{11}\,\hMsun$ (the minimum halo mass in our densest sample) and measure the velocity difference as a function of halo mass. We define the velocity bias as the magnitude of the velocity difference between the host halo and its central subhalo, $\Delta v_{\mathrm{cen}}$. We then compute the ratio of the mean $\Delta v_{\mathrm{cen}}$ between the two simulations and parameterise this ratio as a function of halo mass in \MTNGdmo, which allows us to correct all haloes accordingly.

The algorithm developed to modify the positions and velocities of central and satellite subhaloes can be applied to any dark-matter-only simulation to emulate the baryonic effects observed in \MTNG. These modifications, however, do not constitute a universal prescription for baryonification; they merely reproduce the behaviour specific to this galaxy-formation model. Baryonic effects vary across simulations, and their amplitude in the real Universe remains uncertain (e.g. \citealt{Arico:2020,Arico:2021}; see also Appendix~\ref{sec:zoom} for a discussion of how different galaxy-formation models affect subhalo positions and velocities). Furthermore, although baryonification models typically depend on halo properties such as distance to the halo centre and halo mass (as we do here), secondary dependencies on halo concentration or large-scale environment may also play a role (Wang et al., in prep.). The purpose of this work is not to build a general baryonification model for subhaloes but rather to quantify, to first order, how baryons affect galaxy clustering. Nevertheless, the methodology presented here provides a useful framework for developing more universal subhalo baryonification schemes.

We refer to \MTNGdmoMOD\ as the \MTNGdmo\ simulation with subhalo positions and velocities modified according to the prescriptions above. In Fig.~\ref{Fig:clustering}, we compare the clustering of subhaloes in \MTNGdmo\ and \MTNGdmoMOD. We compute the real-space two-point correlation function ($\xi(r)$), the projected correlation function in redshift space with an integration limit of $30~\hMpc$ ($w_p(r)$), the monopole of the redshift-space correlation function ($\xi_{\ell=0}(r)$), and the quadrupole ($\xi_{\ell=2}(r)$) for the three number-density samples used in this work. All clustering measurements were obtained with \texttt{CORRFUNC} \citep{corrfunc}. To test the validity of this approach, we measure the clustering of the \MTNG\ and the \MTNGdmoMOD\ using the matching catalogues, finding an overall good agreement in the clustering (see Appendix~\ref{sec:Mimicking} for more details).

The real- and redshift-space correlation functions, as well as the projected correlation function, show clustering differences of 10–20\% on small scales, which gradually diminish and become negligible at scales of 1–3~$\hMpc$ (bottom panels of Fig.~\ref{Fig:clustering}). \cite{ChavesMontero:2016} also studied this effect for galaxies selected in bins of stellar mass and found compatible results. As expected from previous results, subhaloes in \MTNGdmoMOD\ are more strongly clustered than in \MTNGdmo. The impact of the velocity correction is modest for the monopole, producing differences similar to those seen in the real-space two-point function. Although clustering amplitudes vary among the number-density samples, the relative ratios—interpreted as the effect of baryons on galaxy clustering—show no significant dependence on number density.

The quadrupole of the correlation function, on the other hand, exhibits a clear dependence on number density: less dense samples (corresponding to more massive or luminous galaxies in an abundance-matching context) show larger amplitude differences than the densest samples. This behaviour, however, partly reflects the choice of presentation. Unlike the other statistics, where we plot ratios, here we show the difference in the signal multiplied by $r^2$, as the quadrupole crosses zero and the ratio would be ill-defined. When inspecting the ratio directly (not shown here), the number-density dependence weakens considerably. In any case, as discussed in the following section, the quadrupole amplitude is the least sensitive of the statistics considered here and does not warrant further interpretation at this stage. We therefore conclude that the effect of baryons on galaxy clustering is similar across all number densities examined in this work.

\subsection{Baryonic effects and selection effects}
\label{sec:sel_effect}

Baryonic effects are not the only systematics that produce differences in the spatial distributions of satellite galaxies within haloes. Selection effects—i.e. the particular criteria used to define a galaxy sample in a survey—also modify the spatial and kinematic properties of satellite populations. For instance, the satellites in a star-forming sample tend to reside in the outer regions of haloes \citep{Orsi:2018}, whereas galaxies selected by stellar mass are predominantly located in the inner regions. To account for both selection and baryonic effects, modern abundance-matching approaches (e.g. \citealt{Contreras:2021_SHAMe, Favole:2022, Rocher:2023, Ortega:2024}) and halo-occupation-based models (e.g. \citealt{Carretero:2015, Yuan:2022}; Gonzalez et al., in prep.) include parameters that explicitly control the spatial and/or velocity distribution of satellite galaxies. This can be implemented either by selecting dark-matter particles or subhaloes in specific regions of the halo (inner or outer) and populating them with galaxies, or by assigning galaxies directly through analytical prescriptions.

To place the impact of selection effects in context with that of baryonic effects, we compare the ratios and differences of several two-point statistics: the real-space and projected correlation functions, and the monopole and quadrupole of the redshift-space correlation function. These quantities are evaluated for galaxies selected by stellar mass, and for those selected by their $r$-band magnitude (\Mr), $U$-band magnitude (\Mu), and star formation rate (SFR). All samples are drawn from the hydrodynamic \MTNG\ simulation and contain galaxies with the highest stellar masses and SFRs, and the lowest magnitudes (i.e. highest luminosities), at a fixed number density of $0.0316~\ihMpcC$ (Fig.~\ref{Fig:SelEff}).

We find that the clustering differences among the selection-based samples are substantially larger than those induced by baryonic effects. On large scales, the projected and real-space correlation functions—both with and without redshift-space distortions—exhibit constant relative offsets. These reflect differences in the large-scale bias of the galaxy samples, which arise from variations in the satellite fraction and in the mass distribution of their host haloes (the dominant effect), as well as from different levels of galaxy assembly bias \citep{Gao:2007, Croton:2007, Zehavi:2018, Artale:2018, Contreras:2019, GarciaMoreno:2025}.

For the real-space and projected correlation functions, we observe strong differences in the one-halo term, which vary in the opposite direction to the shifts produced by baryonic effects. As previously discussed, moving from redder samples (stellar-mass-selected galaxies) to bluer samples (\Mr, \Mu, and SFR) selects galaxies that increasingly occupy the outer regions of haloes, thereby reducing the small-scale clustering amplitude. This trend is not present for the monopole of the correlation function. We interpret this as evidence that selecting outer-halo galaxies also selects systems with different velocity profiles: satellites in the outskirts have lower orbital velocities and follow more coherent orbits around the halo centre. As a result, they appear dynamically closer to the centre than the faster-moving satellites in the inner regions. These velocity-profile differences become more evident in the quadrupole, which shows larger deviations as we move towards bluer samples. The competition between spatial and velocity effects reduces the overall difference in the monopole at the smallest scales.

The differences we find among the various selection functions are far greater than those caused by baryonic effects (indicated by the grey dashed lines in the figures). This is particularly evident for the quadrupole, where the selection-induced deviations exceed the baryonic ones by more than an order of magnitude. We also tested different number densities and enforced equal satellite fractions across samples (not shown), finding some variation in the detailed trends but reaching the same conclusion: selection effects have a significantly stronger impact on galaxy clustering than baryonic effects.

\section{Summary}
\label{sec:conclusions}

In this work, we have examined the impact of baryonic effects on the positions and velocities of satellite galaxies. We used galaxies from \MTNG, one of the largest high-resolution cosmological hydrodynamic simulations available, and matched them to subhaloes from a dark-matter-only simulation, \MTNGdmo, which shares the same initial conditions, cosmology, and resolution as its hydrodynamic counterpart. Satellites in both simulations were matched by tracing their progenitors to the epoch when they were central objects and selecting those with similar positions and halo masses. This procedure ensures a reliable correspondence between simulations, even for satellites that fell into their host haloes several gigayears ago and whose present-day positions and masses differ substantially.

Here we summarise the main results of this work:

\begin{itemize}

 \item Satellite galaxies in the hydrodynamic simulation are, on average, 3–4\% closer to the halo centre than the subhaloes in the dark-matter-only simulation (Fig.~\ref{Fig:mean_dist}). When analysing the radial profile of satellite distribution, most of the difference originates from objects located in the innermost halo regions (100–300~$\hkpc$), where the hydrodynamic simulation exhibits up to a 10\% higher abundance (Fig.~\ref{Fig:prof_dist}). We find no significant dependence on halo mass.

 \item The median satellite velocity dispersion of haloes above $10^{14}\,\hMsun$ is similar in both simulations. For lower-mass haloes, satellites in the dark-matter-only simulation move up to 3\% faster than in the hydrodynamic run (Fig.~\ref{Fig:mean_vel}). When considering the dependence of velocity on distance from the halo centre, galaxies within 100–200~$\hkpc$ show the largest differences, with satellites in the dark-matter-only simulation moving up to 10\% faster than those in the hydrodynamic simulation. These differences show little correlation with halo mass (Fig.~\ref{Fig:prof_vel}). We do not observe a change in the median velocity dispersion for the most massive haloes in the hydrodynamic run, since although satellites move more slowly in the inner regions, their higher abundance there compensates for this effect. Because satellites move faster in the inner halo, the combination of slower velocities and higher density yields a similar overall velocity dispersion.

 \item We developed an analytic approach to modify the positions and velocities of subhaloes in a dark-matter-only simulation so that they mimic the baryonic effects observed in \MTNG\ (Section~\ref{sec:clustering} and Appendix~\ref{sec:Mimicking}). The corrections are parameterised as functions of the satellite distance from the halo centre and host-halo mass.

 \item We compared the subhalo clustering in \MTNGdmo\ with that in \MTNGdmoMOD, a modified version of the dark-matter-only simulation in which subhalo positions and velocities were adjusted to reproduce the baryonic effects seen in the hydrodynamic \MTNG\ run. We did not directly compare \MTNG\ and \MTNGdmo\ because the matching between the two is not fully complete, and differences in selection functions could introduce systematic biases, but we test the validity of our approach by comparing the clustering of the matched catalogues and find good agreement with the \MTNG. We find that \MTNGdmoMOD\ exhibits higher clustering amplitudes in the two-point, projected, and monopole correlation functions—by up to 10–20\% at scales of $\sim100~\hkpc$—with the differences vanishing at scales of $\sim1~\hMpc$. We find no significant dependence on number density for $\vpeak$-selected subhalo samples. The quadrupole shows only minor differences between models, with ${\rm \Delta r^2 \xi_{\ell=2} \approx 2}~h^{-2}\,{\rm, Mpc^2}$ (Fig.~\ref{Fig:clustering}).

 \item We also compared the clustering of galaxies in \MTNG\ selected by different intrinsic properties, including stellar mass, \Mr, \Mu, and SFR (Fig.~\ref{Fig:SelEff}). The clustering differences among these selections are much larger than those caused by baryonic effects. Only the monopole shows differences of comparable magnitude at small scales, likely due to compensating effects between galaxy positions and velocities. More complex colour-based selection functions, such as those used in surveys like DESI or \textit{Euclid} (e.g. \citealt{DESI_color1,DESI_color2,EUCLID_color}), may produce even larger differences.
 
\end{itemize}

We emphasise that the baryonic effects quantified here correspond to the fiducial \MTNG\ galaxy-formation model and may not precisely reflect those in the real Universe. In fact, the baryonic suppression of the mass distribution in \MTNG\ is somewhat weaker than that predicted by other cosmological hydrodynamic simulations (e.g. \citealt{Arico:2020,Arico:2021}). In Appendix~\ref{sec:zoom}, we use a suite of 26 zoom-in hydrodynamic simulations with the same initial conditions as \MTNG, each incorporating modified baryonic physics, to assess how such changes affect satellite properties. This set includes 452 re-simulated haloes (29 with $M_{\rm h}>10^{13}\,\hMsun$). Different physical prescriptions yield different amplitudes for baryonic effects, with the most extreme models predicting an increase by up to a factor of two relative to the fiducial case. However, the limited number of massive haloes prevents robust statistical constraints; these results should thus be regarded as indicative, providing an upper bound on the expected amplitude of baryonic effects.

Even if baryonic effects are enhanced by a factor of two or three due to differences in galaxy-formation physics, they remain subdominant compared to selection effects. Nevertheless, they are not negligible, as they correspond to 1-3 $\sigma$ differences in clustering assuming DESI 1\% errors \citep{Yuan:2024}. These discrepancies will be larger for DR1 or other future releases of DESI. Therefore, baryonic effects should be explicitly accounted for in galaxy-modelling frameworks.

Overall, our results provide a detailed view of how baryons influence position and velocities of satellite galaxies and how these effects depend on halo mass and radial position. These findings will contribute to improving how baryonic corrections are implemented in empirical models, particularly in conjunction with selection effects. Although developing a universal baryonification model was not the goal of this paper, the method introduced here offers a promising basis for extending baryonic corrections to any dark-matter-only simulation. In future work, we will investigate in greater detail how these effects depend on galaxy-formation physics and test the performance of empirical models in reproducing them.

\begin{acknowledgements}
We thank Giovanni Arico and Jonas Chaves-Montero for some usfull suggestions and comments.
SC acknowledges the support of the ``Ram\'on y Cajal'' fellowship (RYC2023-043783-I).
REA received support from grant PID2024-161003NB-I00 funded by MICIU/AEI/10.13039/501100011033 and by ERDF/EU.
SB is supported by the UKRI Future Leaders Fellowship [grant numbers MR/V023381/1 and UKRI2044].
FM acknowledges support by the Simons Collaboration on ``Learning the Universe''.
VS acknowledges support by the Deutsche Forschungsgemeinschaft (DFG, German Research Foundation) under Germany’s Excellence Strategy – EXC-2094 – 390783311.
{\it Author Contributions Statement:} The idea for this project was developed by SC, with essential contributions from REA. SC analysed the data, performed the calculations, and wrote the manuscript. REA provided critical comments throughout the development of the project. SB, BH, LH, RP, and VS developed the MTNG simulation. RP provided critical feedback on Appendix A, and VS offered technical assistance during the analysis of the simulations. REA, FM, and VS developed the zoom-in simulations used in Appendix B, and FM ran most of them. All authors contributed comments on the manuscript. From the third author onward, the author list is ordered alphabetically.
\end{acknowledgements}
\bibliographystyle{aa} % style aa.bst
\bibliography{aa.bib} % your references Yourfile.bib

@ARTICLE{Vogelsberger:2014a,
       author = {{Vogelsberger}, Mark and {Genel}, Shy and {Springel}, Volker and {Torrey}, Paul and {Sijacki}, Debora and {Xu}, Dandan and {Snyder}, Greg and {Nelson}, Dylan and {Hernquist}, Lars},
        title = "{Introducing the Illustris Project: simulating the coevolution of dark and visible matter in the Universe}",
      journal = {\mnras},
         year = 2014,
        month = oct,
       volume = {444},
       number = {2},
        pages = {1518-1547},
          doi = {10.1093/mnras/stu1536},
archivePrefix = {arXiv},
       eprint = {1405.2921},
 primaryClass = {astro-ph.CO},
       adsurl = {https://ui.adsabs.harvard.edu/abs/2014MNRAS.444.1518V}
}

@ARTICLE{Vogelsberger:2014b,
       author = {{Vogelsberger}, M. and {Genel}, S. and {Springel}, V. and {Torrey}, P. and {Sijacki}, D. and {Xu}, D. and {Snyder}, G. and {Bird}, S. and {Nelson}, D. and {Hernquist}, L.},
        title = "{Properties of galaxies reproduced by a hydrodynamic simulation}",
      journal = {\nat},
         year = 2014,
        month = may,
       volume = {509},
       number = {7499},
        pages = {177-182},
          doi = {10.1038/nature13316},
archivePrefix = {arXiv},
       eprint = {1405.1418},
 primaryClass = {astro-ph.CO},
       adsurl = {https://ui.adsabs.harvard.edu/abs/2014Natur.509..177V}
}

@ARTICLE{DOnghia:2010,
       author = {{D'Onghia}, Elena and {Springel}, Volker and {Hernquist}, Lars and {Keres}, Dusan},
        title = "{Substructure Depletion in the Milky Way Halo by the Disk}",
      journal = {\apj},
         year = 2010,
        month = feb,
       volume = {709},
       number = {2},
        pages = {1138-1147},
          doi = {10.1088/0004-637X/709/2/1138},
archivePrefix = {arXiv},
       eprint = {0907.3482},
 primaryClass = {astro-ph.CO},
       adsurl = {https://ui.adsabs.harvard.edu/abs/2010ApJ...709.1138D}
}

@ARTICLE{GarciaMoreno:2025,
       author = {{Garc{\'\i}a-Moreno}, Sergio and {Chaves-Montero}, Jon{\'a}s},
        title = "{Measuring and predicting galaxy assembly bias across galaxy samples}",
      journal = {arXiv e-prints},
         year = 2025,
        month = apr,
          eid = {arXiv:2504.06770},
        pages = {arXiv:2504.06770},
          doi = {10.48550/arXiv.2504.06770},
archivePrefix = {arXiv},
       eprint = {2504.06770},
 primaryClass = {astro-ph.CO},
       adsurl = {https://ui.adsabs.harvard.edu/abs/2025arXiv250406770G}
}

@ARTICLE{Burger:2025,
       author = {{Burger}, Jan D. and {Springel}, Volker and {Ostriker}, Eve C. and {Kim}, Chang-Goo and {Jeffreson}, Sarah M.~R. and {Smith}, Matthew C. and {Pakmor}, R{\"u}diger and {Hassan}, Sultan and {Fielding}, Drummond and {Hernquist}, Lars and {Bryan}, Greg L. and {Somerville}, Rachel S. and {Bennett}, Jake S. and {Weinberger}, Rainer},
        title = "{Applying a star formation model calibrated on high-resolution interstellar medium simulations to cosmological simulations of galaxy formation}",
      journal = {\mnras},
         year = 2025,
        month = dec,
       volume = {544},
       number = {2},
        pages = {1390-1411},
          doi = {10.1093/mnras/staf1720},
archivePrefix = {arXiv},
       eprint = {2502.13244},
 primaryClass = {astro-ph.GA},
       adsurl = {https://ui.adsabs.harvard.edu/abs/2025MNRAS.544.1390B}
}

@ARTICLE{Miller:2020,
       author = {{Miller}, Tim B. and {van den Bosch}, Frank C. and {Green}, Sheridan B. and {Ogiya}, Go},
        title = "{Dynamical self-friction: how mass loss slows you down}",
      journal = {\mnras},
         year = 2020,
        month = jul,
       volume = {495},
       number = {4},
        pages = {4496-4507},
          doi = {10.1093/mnras/staa1450},
archivePrefix = {arXiv},
       eprint = {2001.06489},
 primaryClass = {astro-ph.GA},
       adsurl = {https://ui.adsabs.harvard.edu/abs/2020MNRAS.495.4496M}
}

@ARTICLE{Ostriker:1999,
       author = {{Ostriker}, Eve C.},
        title = "{Dynamical Friction in a Gaseous Medium}",
      journal = {\apj},
         year = 1999,
        month = mar,
       volume = {513},
       number = {1},
        pages = {252-258},
          doi = {10.1086/306858},
archivePrefix = {arXiv},
       eprint = {astro-ph/9810324},
 primaryClass = {astro-ph},
       adsurl = {https://ui.adsabs.harvard.edu/abs/1999ApJ...513..252O}
}

@ARTICLE{Ogiya:2016,
       author = {{Ogiya}, Go and {Burkert}, Andreas},
        title = "{Dynamical friction and scratches of orbiting satellite galaxies on host systems}",
      journal = {\mnras},
         year = 2016,
        month = apr,
       volume = {457},
       number = {2},
        pages = {2164-2172},
          doi = {10.1093/mnras/stw091},
archivePrefix = {arXiv},
       eprint = {1510.07892},
 primaryClass = {astro-ph.GA},
       adsurl = {https://ui.adsabs.harvard.edu/abs/2016MNRAS.457.2164O}
}

@ARTICLE{Yuan:2024,
       author = {{Yuan}, Sihan and {Zhang}, Hanyu and {Ross}, Ashley J. and {Donald-McCann}, Jamie and {Hadzhiyska}, Boryana and {Wechsler}, Risa H. and {Zheng}, Zheng and {Alam}, Shadab and {Gonzalez-Perez}, Violeta and {Aguilar}, Jessica Nicole and {Ahlen}, Steven and {Bianchi}, Davide and {Brooks}, David and {de la Macorra}, Axel and {Fanning}, Kevin and {Forero-Romero}, Jaime E. and {Honscheid}, Klaus and {Ishak}, Mustapha and {Kehoe}, Robert and {Lasker}, James and {Landriau}, Martin and {Manera}, Marc and {Martini}, Paul and {Meisner}, Aaron and {Miquel}, Ramon and {Moustakas}, John and {Nadathur}, Seshadri and {Newman}, Jeffrey A. and {Nie}, Jundan and {Percival}, Will and {Poppett}, Claire and {Rocher}, Antoine and {Rossi}, Graziano and {Sanchez}, Eusebio and {Samushia}, Lado and {Schubnell}, Michael and {Seo}, Hee-Jong and {Tarl{\'e}}, Gregory and {Weaver}, Benjamin Alan and {Yu}, Jiaxi and {Zhou}, Zhimin and {Zou}, Hu},
        title = "{The DESI one-per cent survey: exploring the halo occupation distribution of luminous red galaxies and quasi-stellar objects with ABACUSSUMMIT}",
      journal = {\mnras},
         year = 2024,
        month = may,
       volume = {530},
       number = {1},
        pages = {947-965},
          doi = {10.1093/mnras/stae359},
archivePrefix = {arXiv},
       eprint = {2306.06314},
 primaryClass = {astro-ph.CO},
       adsurl = {https://ui.adsabs.harvard.edu/abs/2024MNRAS.530..947Y}
}

@ARTICLE{Akino:2022,
       author = {{Akino}, Daichi and {Eckert}, Dominique and {Okabe}, Nobuhiro and {Sereno}, Mauro and {Umetsu}, Keiichi and {Oguri}, Masamune and {Gastaldello}, Fabio and {Chiu}, I.-Non and {Ettori}, Stefano and {Evrard}, August E. and {Farahi}, Arya and {Maughan}, Ben and {Pierre}, Marguerite and {Ricci}, Marina and {Valtchanov}, Ivan and {McCarthy}, Ian and {McGee}, Sean and {Miyazaki}, Satoshi and {Nishizawa}, Atsushi J. and {Tanaka}, Masayuki},
        title = "{HSC-XXL: Baryon budget of the 136 XXL groups and clusters}",
      journal = {\pasj},
         year = 2022,
        month = feb,
       volume = {74},
       number = {1},
        pages = {175-208},
          doi = {10.1093/pasj/psab115},
archivePrefix = {arXiv},
       eprint = {2111.10080},
 primaryClass = {astro-ph.CO},
       adsurl = {https://ui.adsabs.harvard.edu/abs/2022PASJ...74..175A}
}

@ARTICLE{DESI_color2,
       author = {{Myers}, Adam D. and {Moustakas}, John and {Bailey}, Stephen and {Weaver}, Benjamin A. and {Cooper}, Andrew P. and {Forero-Romero}, Jaime E. and {Abolfathi}, Bela and {Alexander}, David M. and {Brooks}, David and {Chaussidon}, Edmond and {Chuang}, Chia-Hsun and {Dawson}, Kyle and {Dey}, Arjun and {Dey}, Biprateep and {Dhungana}, Govinda and {Doel}, Peter and {Fanning}, Kevin and {Gazta{\~n}aga}, Enrique and {Gontcho A Gontcho}, Satya and {Gonzalez-Morales}, Alma X. and {Hahn}, ChangHoon and {Herrera-Alcantar}, Hiram K. and {Honscheid}, Klaus and {Ishak}, Mustapha and {Karim}, Tanveer and {Kirkby}, David and {Kisner}, Theodore and {Koposov}, Sergey E. and {Kremin}, Anthony and {Lan}, Ting-Wen and {Landriau}, Martin and {Lang}, Dustin and {Levi}, Michael E. and {Magneville}, Christophe and {Napolitano}, Lucas and {Martini}, Paul and {Meisner}, Aaron and {Newman}, Jeffrey A. and {Palanque-Delabrouille}, Nathalie and {Percival}, Will and {Poppett}, Claire and {Prada}, Francisco and {Raichoor}, Anand and {Ross}, Ashley J. and {Schlafly}, Edward F. and {Schlegel}, David and {Schubnell}, Michael and {Tan}, Ting and {Tarle}, Gregory and {Wilson}, Michael J. and {Y{\`e}che}, Christophe and {Zhou}, Rongpu and {Zhou}, Zhimin and {Zou}, Hu},
        title = "{The Target-selection Pipeline for the Dark Energy Spectroscopic Instrument}",
      journal = {\aj},
         year = 2023,
        month = feb,
       volume = {165},
       number = {2},
          eid = {50},
        pages = {50},
          doi = {10.3847/1538-3881/aca5f9},
archivePrefix = {arXiv},
       eprint = {2208.08518},
 primaryClass = {astro-ph.IM},
       adsurl = {https://ui.adsabs.harvard.edu/abs/2023AJ....165...50M}
}

@ARTICLE{DESI_color1,
       author = {{Raichoor}, A. and {Moustakas}, J. and {Newman}, Jeffrey A. and {Karim}, T. and {Ahlen}, S. and {Alam}, Shadab and {Bailey}, S. and {Brooks}, D. and {Dawson}, K. and {de la Macorra}, A. and {de Mattia}, A. and {Dey}, A. and {Dey}, Biprateep and {Dhungana}, G. and {Eftekharzadeh}, S. and {Eisenstein}, D.~J. and {Fanning}, K. and {Font-Ribera}, A. and {Garc{\'\i}a-Bellido}, J. and {Gazta{\~n}aga}, E. and {A Gontcho}, S. Gontcho and {Guy}, J. and {Honscheid}, K. and {Ishak}, M. and {Kehoe}, R. and {Kisner}, T. and {Kremin}, Anthony and {Lan}, Ting-Wen and {Landriau}, M. and {Le Guillou}, L. and {Levi}, Michael E. and {Magneville}, C. and {Manera}, M. and {Martini}, P. and {Meisner}, Aaron M. and {Myers}, Adam D. and {Nie}, Jundan and {Palanque-Delabrouille}, N. and {Percival}, W.~J. and {Poppett}, C. and {Prada}, F. and {Ross}, A.~J. and {Ruhlmann-Kleider}, V. and {Sabiu}, C.~G. and {Schlafly}, E.~F. and {Schlegel}, D. and {Tarl{\'e}}, Gregory and {Weaver}, B.~A. and {Y{\`e}che}, Christophe and {Zhou}, Rongpu and {Zhou}, Zhimin and {Zou}, H.},
        title = "{Target Selection and Validation of DESI Emission Line Galaxies}",
      journal = {\aj},
         year = 2023,
        month = mar,
       volume = {165},
       number = {3},
          eid = {126},
        pages = {126},
          doi = {10.3847/1538-3881/acb213},
archivePrefix = {arXiv},
       eprint = {2208.08513},
 primaryClass = {astro-ph.CO},
       adsurl = {https://ui.adsabs.harvard.edu/abs/2023AJ....165..126R}
}

@ARTICLE{Euclid_color,
       author = {{Euclid Collaboration} and {Lesci}, G.~F. and {Sereno}, M. and {Radovich}, M. and {Castignani}, G. and {Bisigello}, L. and {Marulli}, F. and {Moscardini}, L. and {Baumont}, L. and {Covone}, G. and {Farrens}, S. and {Giocoli}, C. and {Ingoglia}, L. and {Miranda La Hera}, S. and {Vannier}, M. and {Biviano}, A. and {Maurogordato}, S. and {Aghanim}, N. and {Amara}, A. and {Andreon}, S. and {Auricchio}, N. and {Baldi}, M. and {Bardelli}, S. and {Bender}, R. and {Bodendorf}, C. and {Bonino}, D. and {Branchini}, E. and {Brescia}, M. and {Brinchmann}, J. and {Camera}, S. and {Capobianco}, V. and {Carbone}, C. and {Carretero}, J. and {Casas}, S. and {Castander}, F.~J. and {Castellano}, M. and {Cavuoti}, S. and {Cimatti}, A. and {Congedo}, G. and {Conselice}, C.~J. and {Conversi}, L. and {Copin}, Y. and {Corcione}, L. and {Courbin}, F. and {Courtois}, H.~M. and {Da Silva}, A. and {Degaudenzi}, H. and {Di Giorgio}, A.~M. and {Dinis}, J. and {Dubath}, F. and {Duncan}, C.~A.~J. and {Dupac}, X. and {Dusini}, S. and {Farina}, M. and {Ferriol}, S. and {Fosalba}, P. and {Fotopoulou}, S. and {Frailis}, M. and {Franceschi}, E. and {Franzetti}, P. and {Fumana}, M. and {Galeotta}, S. and {Garilli}, B. and {Gillis}, B. and {Grazian}, A. and {Grupp}, F. and {Haugan}, S.~V.~H. and {Hook}, I. and {Hormuth}, F. and {Hornstrup}, A. and {Hudelot}, P. and {Jahnke}, K. and {K{\"u}mmel}, M. and {Kermiche}, S. and {Kiessling}, A. and {Kilbinger}, M. and {Kubik}, B. and {Kunz}, M. and {Kurki-Suonio}, H. and {Ligori}, S. and {Lilje}, P.~B. and {Lindholm}, V. and {Lloro}, I. and {Maiorano}, E. and {Mansutti}, O. and {Marggraf}, O. and {Markovic}, K. and {Martinet}, N. and {Massey}, R. and {Medinaceli}, E. and {Melchior}, M. and {Mellier}, Y. and {Meneghetti}, M. and {Merlin}, E. and {Meylan}, G. and {Moresco}, M. and {Munari}, E. and {Nakajima}, R. and {Niemi}, S.-M. and {Padilla}, C. and {Paltani}, S. and {Pasian}, F. and {Pedersen}, K. and {Pettorino}, V. and {Pires}, S. and {Polenta}, G. and {Poncet}, M. and {Popa}, L.~A. and {Pozzetti}, L. and {Raison}, F. and {Rebolo}, R. and {Renzi}, A. and {Rhodes}, J. and {Riccio}, G. and {Romelli}, E. and {Roncarelli}, M. and {Rossetti}, E. and {Saglia}, R. and {Sapone}, D. and {Sartoris}, B. and {Schirmer}, M. and {Schneider}, P. and {Secroun}, A. and {Seidel}, G. and {Serrano}, S. and {Sirignano}, C. and {Sirri}, G. and {Skottfelt}, J. and {Stanco}, L. and {Starck}, J.-L. and {Tallada-Cresp{\'\i}}, P. and {Taylor}, A.~N. and {Teplitz}, H.~I. and {Tereno}, I. and {Toledo-Moreo}, R. and {Torradeflot}, F. and {Tutusaus}, I. and {Valentijn}, E.~A. and {Valenziano}, L. and {Vassallo}, T. and {Veropalumbo}, A. and {Wang}, Y. and {Weller}, J. and {Zacchei}, A. and {Zamorani}, G. and {Zoubian}, J. and {Zucca}, E. and {Bolzonella}, M. and {Bozzo}, E. and {Colodro-Conde}, C. and {Di Ferdinando}, D. and {Graci{\'a}-Carpio}, J. and {Marcin}, S. and {Mauri}, N. and {Neissner}, C. and {Nucita}, A.~A. and {Sakr}, Z. and {Scottez}, V. and {Tenti}, M. and {Viel}, M. and {Wiesmann}, M. and {Akrami}, Y. and {Anselmi}, S. and {Baccigalupi}, C. and {Ballardini}, M. and {Borgani}, S. and {Borlaff}, A.~S. and {Bruton}, S. and {Burigana}, C. and {Cabanac}, R. and {Calabro}, A. and {Cappi}, A. and {Carvalho}, C.~S. and {Castro}, T. and {Ca{\~n}as-Herrera}, G. and {Chambers}, K.~C. and {Cooray}, A.~R. and {Coupon}, J. and {Cucciati}, O. and {Davini}, S. and {de la Torre}, S. and {De Lucia}, G. and {Desprez}, G. and {Di Domizio}, S. and {Dole}, H. and {D{\'\i}az-S{\'a}nchez}, A. and {Escartin Vigo}, J.~A. and {Escoffier}, S. and {Ferrero}, I. and {Finelli}, F. and {Gabarra}, L. and {Ganga}, K. and {Garc{\'\i}a-Bellido}, J. and {Giacomini}, F. and {Gozaliasl}, G. and {Gwyn}, S. and {Hildebrandt}, H. and {Huertas-Company}, M. and {Jimenez Mu{\~n}oz}, A. and {Kajava}, J.~J.~E.},
        title = "{Euclid preparation. XXXVII. Galaxy colour selections with Euclid and ground photometry for cluster weak-lensing analyses}",
      journal = {\aap},
         year = 2024,
        month = apr,
       volume = {684},
          eid = {A139},
        pages = {A139},
          doi = {10.1051/0004-6361/202348743},
archivePrefix = {arXiv},
       eprint = {2311.16239},
 primaryClass = {astro-ph.CO},
       adsurl = {https://ui.adsabs.harvard.edu/abs/2024A&A...684A.139E}
}

@ARTICLE{Carretero:2015,
       author = {{Carretero}, J. and {Castander}, F.~J. and {Gazta{\~n}aga}, E. and {Crocce}, M. and {Fosalba}, P.},
        title = "{An algorithm to build mock galaxy catalogues using MICE simulations}",
      journal = {\mnras},
         year = 2015,
        month = feb,
       volume = {447},
       number = {1},
        pages = {646-670},
          doi = {10.1093/mnras/stu2402},
archivePrefix = {arXiv},
       eprint = {1411.3286},
 primaryClass = {astro-ph.GA},
       adsurl = {https://ui.adsabs.harvard.edu/abs/2015MNRAS.447..646C}
}

@ARTICLE{Artale:2018,
       author = {{Artale}, M. Celeste and {Zehavi}, Idit and {Contreras}, Sergio and
         {Norberg}, Peder},
        title = "{The impact of assembly bias on the halo occupation in hydrodynamical simulations}",
      journal = {\mnras},
         year = "2018",
        month = "Nov",
       volume = {480},
       number = {3},
        pages = {3978-3992},
          doi = {10.1093/mnras/sty2110},
archivePrefix = {arXiv},
       eprint = {1805.06938},
 primaryClass = {astro-ph.GA},
       adsurl = {https://ui.adsabs.harvard.edu/abs/2018MNRAS.480.3978A}
}

@ARTICLE{Zehavi:2018,
       author = {{Zehavi}, Idit and {Contreras}, Sergio and {Padilla}, Nelson and {Smith}, Nicholas J. and {Baugh}, Carlton M. and {Norberg}, Peder},
        title = "{The Impact of Assembly Bias on the Galaxy Content of Dark Matter Halos}",
      journal = {\apj},
         year = 2018,
        month = jan,
       volume = {853},
       number = {1},
          eid = {84},
        pages = {84},
          doi = {10.3847/1538-4357/aaa54a},
archivePrefix = {arXiv},
       eprint = {1706.07871},
 primaryClass = {astro-ph.GA},
       adsurl = {https://ui.adsabs.harvard.edu/abs/2018ApJ...853...84Z}
}

@ARTICLE{Gao:2007,
   author = {{Gao}, L. and {White}, S.~D.~M.},
    title = "{Assembly bias in the clustering of dark matter haloes}",
  journal = {\mnras},
   eprint = {astro-ph/0611921},
     year = 2007,
    month = apr,
   volume = 377,
    pages = {L5-L9},
      doi = {10.1111/j.1745-3933.2007.00292.x},
   adsurl = {http://adsabs.harvard.edu/abs/2007MNRAS.377L...5G}
}

@ARTICLE{Croton:2007,
   author = {{Croton}, D.~J. and {Gao}, L. and {White}, S.~D.~M.},
    title = "{Halo assembly bias and its effects on galaxy clustering}",
  journal = {\mnras},
   eprint = {astro-ph/0605636},
     year = 2007,
    month = feb,
   volume = 374,
    pages = {1303-1309},
      doi = {10.1111/j.1365-2966.2006.11230.x},
   adsurl = {http://adsabs.harvard.edu/abs/2007MNRAS.374.1303C}
}

@ARTICLE{Yuan:2022,
       author = {{Yuan}, Sihan and {Garrison}, Lehman H. and {Hadzhiyska}, Boryana and {Bose}, Sownak and {Eisenstein}, Daniel J.},
        title = "{ABACUSHOD: a highly efficient extended multitracer HOD framework and its application to BOSS and eBOSS data}",
      journal = {\mnras},
         year = 2022,
        month = mar,
       volume = {510},
       number = {3},
        pages = {3301-3320},
          doi = {10.1093/mnras/stab3355},
archivePrefix = {arXiv},
       eprint = {2110.11412},
 primaryClass = {astro-ph.CO},
       adsurl = {https://ui.adsabs.harvard.edu/abs/2022MNRAS.510.3301Y}
}

@ARTICLE{Favole:2022,
       author = {{Favole}, Ginevra and {Montero-Dorta}, Antonio D. and {Artale}, M. Celeste and {Contreras}, Sergio and {Zehavi}, Idit and {Xu}, Xiaoju},
        title = "{Subhalo abundance matching through the lens of a hydrodynamical simulation}",
      journal = {\mnras},
         year = 2022,
        month = jan,
       volume = {509},
       number = {2},
        pages = {1614-1625},
          doi = {10.1093/mnras/stab3006},
archivePrefix = {arXiv},
       eprint = {2101.10733},
 primaryClass = {astro-ph.GA},
       adsurl = {https://ui.adsabs.harvard.edu/abs/2022MNRAS.509.1614F}
}

@INPROCEEDINGS{corrfunc,
       author = {{Sinha}, Manodeep and {Garrison}, Lehman H.},
        title = "{Corrfunc: Blazing fast correlation functions with AVX512F SIMD Intrinsics}",
    booktitle = {Software Challenges to Exascale Computing. Second Workshop},
         year = 2019,
        month = jan,
        pages = {3-20},
          doi = {10.1007/978-981-13-7729-7_1},
archivePrefix = {arXiv},
       eprint = {1911.08275},
 primaryClass = {astro-ph.IM},
       adsurl = {https://ui.adsabs.harvard.edu/abs/2019scec.conf....3S}
}

@ARTICLE{Arico:2020,
       author = {{Aric{\`o}}, Giovanni and {Angulo}, Raul E. and {Hern{\'a}ndez-Monteagudo}, Carlos and {Contreras}, Sergio and {Zennaro}, Matteo and {Pellejero-Iba{\~n}ez}, Marcos and {Rosas-Guevara}, Yetli},
        title = "{Modelling the large-scale mass density field of the universe as a function of cosmology and baryonic physics}",
      journal = {\mnras},
         year = 2020,
        month = jul,
       volume = {495},
       number = {4},
        pages = {4800-4819},
          doi = {10.1093/mnras/staa1478},
archivePrefix = {arXiv},
       eprint = {1911.08471},
 primaryClass = {astro-ph.CO},
       adsurl = {https://ui.adsabs.harvard.edu/abs/2020MNRAS.495.4800A}
}

@ARTICLE{Arico:2021,
       author = {{Aric{\`o}}, Giovanni and {Angulo}, Raul E. and {Hern{\'a}ndez-Monteagudo}, Carlos and {Contreras}, Sergio and {Zennaro}, Matteo},
        title = "{Simultaneous modelling of matter power spectrum and bispectrum in the presence of baryons}",
      journal = {\mnras},
         year = 2021,
        month = may,
       volume = {503},
       number = {3},
        pages = {3596-3609},
          doi = {10.1093/mnras/stab699},
archivePrefix = {arXiv},
       eprint = {2009.14225},
 primaryClass = {astro-ph.CO},
       adsurl = {https://ui.adsabs.harvard.edu/abs/2021MNRAS.503.3596A}
}

@ARTICLE{Guo:2015,
       author = {{Guo}, Hong and {Zheng}, Zheng and {Zehavi}, Idit and {Behroozi}, Peter S. and {Chuang}, Chia-Hsun and {Comparat}, Johan and {Favole}, Ginevra and {Gottloeber}, Stefan and {Klypin}, Anatoly and {Prada}, Francisco and {Weinberg}, David H. and {Yepes}, Gustavo},
        title = "{Redshift-space clustering of SDSS galaxies - luminosity dependence, halo occupation distribution, and velocity bias}",
      journal = {\mnras},
         year = 2015,
        month = nov,
       volume = {453},
       number = {4},
        pages = {4368-4383},
          doi = {10.1093/mnras/stv1966},
archivePrefix = {arXiv},
       eprint = {1505.07861},
 primaryClass = {astro-ph.CO},
       adsurl = {https://ui.adsabs.harvard.edu/abs/2015MNRAS.453.4368G}
}

@ARTICLE{Planck:2015,
       author = {{Planck Collaboration} and {Ade}, P.~A.~R. and {Aghanim}, N. and
         {Arnaud}, M. and {Ashdown}, M. and {Aumont}, J. and {Baccigalupi}, C. and
         {Banday}, A.~J. and {Barreiro}, R.~B. and {Bartlett}, J.~G. and
         {Bartolo}, N. and {Battaner}, E. and {Battye}, R. and {Benabed}, K. and
         {Beno{\^\i}t}, A. and {Benoit-L{\'e}vy}, A. and {Bernard}, J. -P. and
         {Bersanelli}, M. and {Bielewicz}, P. and {Bock}, J.~J. and
         {Bonaldi}, A. and {Bonavera}, L. and {Bond}, J.~R. and {Borrill}, J. and
         {Bouchet}, F.~R. and {Boulanger}, F. and {Bucher}, M. and
         {Burigana}, C. and {Butler}, R.~C. and {Calabrese}, E. and
         {Cardoso}, J. -F. and {Catalano}, A. and {Challinor}, A. and
         {Chamballu}, A. and {Chary}, R. -R. and {Chiang}, H.~C. and
         {Chluba}, J. and {Christensen}, P.~R. and {Church}, S. and
         {Clements}, D.~L. and {Colombi}, S. and {Colombo}, L.~P.~L. and
         {Combet}, C. and {Coulais}, A. and {Crill}, B.~P. and {Curto}, A. and
         {Cuttaia}, F. and {Danese}, L. and {Davies}, R.~D. and {Davis}, R.~J. and
         {de Bernardis}, P. and {de Rosa}, A. and {de Zotti}, G. and
         {Delabrouille}, J. and {D{\'e}sert}, F. -X. and {Di Valentino}, E. and
         {Dickinson}, C. and {Diego}, J.~M. and {Dolag}, K. and {Dole}, H. and
         {Donzelli}, S. and {Dor{\'e}}, O. and {Douspis}, M. and {Ducout}, A. and
         {Dunkley}, J. and {Dupac}, X. and {Efstathiou}, G. and {Elsner}, F. and
         {En{\ss}lin}, T.~A. and {Eriksen}, H.~K. and {Farhang}, M. and
         {Fergusson}, J. and {Finelli}, F. and {Forni}, O. and {Frailis}, M. and
         {Fraisse}, A.~A. and {Franceschi}, E. and {Frejsel}, A. and
         {Galeotta}, S. and {Galli}, S. and {Ganga}, K. and {Gauthier}, C. and
         {Gerbino}, M. and {Ghosh}, T. and {Giard}, M. and
         {Giraud-H{\'e}raud}, Y. and {Giusarma}, E. and {Gjerl{\o}w}, E. and
         {Gonz{\'a}lez-Nuevo}, J. and {G{\'o}rski}, K.~M. and {Gratton}, S. and
         {Gregorio}, A. and {Gruppuso}, A. and {Gudmundsson}, J.~E. and
         {Hamann}, J. and {Hansen}, F.~K. and {Hanson}, D. and
         {Harrison}, D.~L. and {Helou}, G. and {Henrot-Versill{\'e}}, S. and
         {Hern{\'a}ndez-Monteagudo}, C. and {Herranz}, D. and {Hildebrand
        t}, S.~R. and {Hivon}, E. and {Hobson}, M. and {Holmes}, W.~A. and
         {Hornstrup}, A. and {Hovest}, W. and {Huang}, Z. and
         {Huffenberger}, K.~M. and {Hurier}, G. and {Jaffe}, A.~H. and
         {Jaffe}, T.~R. and {Jones}, W.~C. and {Juvela}, M. and
         {Keih{\"a}nen}, E. and {Keskitalo}, R. and {Kisner}, T.~S. and
         {Kneissl}, R. and {Knoche}, J. and {Knox}, L. and {Kunz}, M. and
         {Kurki-Suonio}, H. and {Lagache}, G. and {L{\"a}hteenm{\"a}ki}, A. and
         {Lamarre}, J. -M. and {Lasenby}, A. and {Lattanzi}, M. and
         {Lawrence}, C.~R. and {Leahy}, J.~P. and {Leonardi}, R. and
         {Lesgourgues}, J. and {Levrier}, F. and {Lewis}, A. and {Liguori}, M. and
         {Lilje}, P.~B. and {Linden-V{\o}rnle}, M. and {L{\'o}pez-Caniego}, M. and
         {Lubin}, P.~M. and {Mac{\'\i}as-P{\'e}rez}, J.~F. and {Maggio}, G. and
         {Maino}, D. and {Mandolesi}, N. and {Mangilli}, A. and {Marchini}, A. and
         {Maris}, M. and {Martin}, P.~G. and {Martinelli}, M. and
         {Mart{\'\i}nez-Gonz{\'a}lez}, E. and {Masi}, S. and {Matarrese}, S. and
         {McGehee}, P. and {Meinhold}, P.~R. and {Melchiorri}, A. and
         {Melin}, J. -B. and {Mendes}, L. and {Mennella}, A. and
         {Migliaccio}, M. and {Millea}, M. and {Mitra}, S. and
         {Miville-Desch{\^e}nes}, M. -A. and {Moneti}, A. and {Montier}, L. and
         {Morgante}, G. and {Mortlock}, D. and {Moss}, A. and {Munshi}, D. and
         {Murphy}, J.~A. and {Naselsky}, P. and {Nati}, F. and {Natoli}, P. and
         {Netterfield}, C.~B. and {N{\o}rgaard-Nielsen}, H.~U. and
         {Noviello}, F. and {Novikov}, D. and {Novikov}, I. and
         {Oxborrow}, C.~A. and {Paci}, F. and {Pagano}, L. and {Pajot}, F. and
         {Paladini}, R. and {Paoletti}, D. and {Partridge}, B. and {Pasian}, F. and
         {Patanchon}, G. and {Pearson}, T.~J. and {Perdereau}, O. and
         {Perotto}, L. and {Perrotta}, F. and {Pettorino}, V. and
         {Piacentini}, F. and {Piat}, M. and {Pierpaoli}, E. and
         {Pietrobon}, D. and {Plaszczynski}, S. and {Pointecouteau}, E. and
         {Polenta}, G. and {Popa}, L. and {Pratt}, G.~W. and {Pr{\'e}zeau}, G. and
         {Prunet}, S. and {Puget}, J. -L. and {Rachen}, J.~P. and
         {Reach}, W.~T. and {Rebolo}, R. and {Reinecke}, M. and
         {Remazeilles}, M. and {Renault}, C. and {Renzi}, A. and
         {Ristorcelli}, I. and {Rocha}, G. and {Rosset}, C. and {Rossetti}, M. and
         {Roudier}, G. and {Rouill{\'e} d'Orfeuil}, B. and {Rowan-Robinson}, M. and
         {Rubi{\~n}o-Mart{\'\i}n}, J.~A. and {Rusholme}, B. and {Said}, N. and
         {Salvatelli}, V. and {Salvati}, L. and {Sandri}, M. and {Santos}, D. and
         {Savelainen}, M. and {Savini}, G. and {Scott}, D. and
         {Seiffert}, M.~D. and {Serra}, P. and {Shellard}, E.~P.~S. and
         {Spencer}, L.~D. and {Spinelli}, M. and {Stolyarov}, V. and
         {Stompor}, R. and {Sudiwala}, R. and {Sunyaev}, R. and {Sutton}, D. and
         {Suur-Uski}, A. -S. and {Sygnet}, J. -F. and {Tauber}, J.~A. and
         {Terenzi}, L. and {Toffolatti}, L. and {Tomasi}, M. and {Tristram}, M. and
         {Trombetti}, T. and {Tucci}, M. and {Tuovinen}, J. and
         {T{\"u}rler}, M. and {Umana}, G. and {Valenziano}, L. and
         {Valiviita}, J. and {Van Tent}, F. and {Vielva}, P. and {Villa}, F. and
         {Wade}, L.~A. and {Wandelt}, B.~D. and {Wehus}, I.~K. and {White}, M. and
         {White}, S.~D.~M. and {Wilkinson}, A. and {Yvon}, D. and {Zacchei}, A. and
         {Zonca}, A.},
        title = "{Planck 2015 results. XIII. Cosmological parameters}",
      journal = {\aap},
         year = 2016,
        month = sep,
       volume = {594},
          eid = {A13},
        pages = {A13},
          doi = {10.1051/0004-6361/201525830},
archivePrefix = {arXiv},
       eprint = {1502.01589},
 primaryClass = {astro-ph.CO},
       adsurl = {https://ui.adsabs.harvard.edu/abs/2016A&A...594A..13P}
}

@ARTICLE{Springel:2005,
   author = {{Springel}, V. and {White}, S.~D.~M. and {Jenkins}, A. and {Frenk}, C.~S. and {Yoshida}, N. and {Gao}, L. and {Navarro}, J. and {Thacker}, R. and 
	{Croton}, D. and {Helly}, J. and {Peacock}, J.~A. and {Cole}, S. et al.},
    title = "{Simulations of the formation, evolution and clustering of galaxies and quasars}",
  journal = {\nat},
   eprint = {arXiv:astro-ph/0504097},
     year = 2005,
    month = jun,
   volume = 435,
    pages = {629-636},
      doi = {10.1038/nature03597},
   adsurl = {http://adsabs.harvard.edu/abs/2005Natur.435..629S}
}

@ARTICLE{AREPO,
       author = {{Springel}, Volker},
        title = "{E pur si muove: Galilean-invariant cosmological hydrodynamical simulations on a moving mesh}",
      journal = {\mnras},
         year = 2010,
        month = jan,
       volume = {401},
       number = {2},
        pages = {791-851},
          doi = {10.1111/j.1365-2966.2009.15715.x},
archivePrefix = {arXiv},
       eprint = {0901.4107},
 primaryClass = {astro-ph.CO},
       adsurl = {https://ui.adsabs.harvard.edu/abs/2010MNRAS.401..791S}
}

@ARTICLE{TNGa,
       author = {{Nelson}, Dylan and {Pillepich}, Annalisa and {Springel}, Volker and
         {Weinberger}, Rainer and {Hernquist}, Lars and {Pakmor}, R{\"u}diger and
         {Genel}, Shy and {Torrey}, Paul and {Vogelsberger}, Mark and
         {Kauffmann}, Guinevere and {Marinacci}, Federico and {Naiman}, Jill},
        title = "{First results from the IllustrisTNG simulations: the galaxy colour bimodality}",
      journal = {\mnras},
         year = 2018,
        month = mar,
       volume = {475},
       number = {1},
        pages = {624-647},
          doi = {10.1093/mnras/stx3040},
archivePrefix = {arXiv},
       eprint = {1707.03395},
 primaryClass = {astro-ph.GA},
       adsurl = {https://ui.adsabs.harvard.edu/abs/2018MNRAS.475..624N}
}

@ARTICLE{TNGb,
       author = {{Springel}, Volker and {Pakmor}, R{\"u}diger and {Pillepich}, Annalisa and
         {Weinberger}, Rainer and {Nelson}, Dylan and {Hernquist}, Lars and
         {Vogelsberger}, Mark and {Genel}, Shy and {Torrey}, Paul and
         {Marinacci}, Federico and {Naiman}, Jill},
        title = "{First results from the IllustrisTNG simulations: matter and galaxy clustering}",
      journal = {\mnras},
         year = 2018,
        month = mar,
       volume = {475},
       number = {1},
        pages = {676-698},
          doi = {10.1093/mnras/stx3304},
archivePrefix = {arXiv},
       eprint = {1707.03397},
 primaryClass = {astro-ph.GA},
       adsurl = {https://ui.adsabs.harvard.edu/abs/2018MNRAS.475..676S}
}

@ARTICLE{TNGc,
       author = {{Marinacci}, Federico and {Vogelsberger}, Mark and
         {Pakmor}, R{\"u}diger and {Torrey}, Paul and {Springel}, Volker and
         {Hernquist}, Lars and {Nelson}, Dylan and {Weinberger}, Rainer and
         {Pillepich}, Annalisa and {Naiman}, Jill and {Genel}, Shy},
        title = "{First results from the IllustrisTNG simulations: radio haloes and magnetic fields}",
      journal = {\mnras},
         year = 2018,
        month = nov,
       volume = {480},
       number = {4},
        pages = {5113-5139},
          doi = {10.1093/mnras/sty2206},
archivePrefix = {arXiv},
       eprint = {1707.03396},
 primaryClass = {astro-ph.CO},
       adsurl = {https://ui.adsabs.harvard.edu/abs/2018MNRAS.480.5113M}
}

@ARTICLE{TNGd,
       author = {{Pillepich}, Annalisa and {Nelson}, Dylan and {Hernquist}, Lars and
         {Springel}, Volker and {Pakmor}, R{\"u}diger and {Torrey}, Paul and
         {Weinberger}, Rainer and {Genel}, Shy and {Naiman}, Jill P. and
         {Marinacci}, Federico and {Vogelsberger}, Mark},
        title = "{First results from the IllustrisTNG simulations: the stellar mass content of groups and clusters of galaxies}",
      journal = {\mnras},
         year = 2018,
        month = mar,
       volume = {475},
       number = {1},
        pages = {648-675},
          doi = {10.1093/mnras/stx3112},
archivePrefix = {arXiv},
       eprint = {1707.03406},
 primaryClass = {astro-ph.GA},
       adsurl = {https://ui.adsabs.harvard.edu/abs/2018MNRAS.475..648P}
}

@ARTICLE{Nelson:2019,
       author = {{Nelson}, Dylan and {Springel}, Volker and {Pillepich}, Annalisa and {Rodriguez-Gomez}, Vicente and {Torrey}, Paul and {Genel}, Shy and {Vogelsberger}, Mark and {Pakmor}, Ruediger and {Marinacci}, Federico and {Weinberger}, Rainer and {Kelley}, Luke and {Lovell}, Mark and {Diemer}, Benedikt and {Hernquist}, Lars},
        title = "{The IllustrisTNG simulations: public data release}",
      journal = {Computational Astrophysics and Cosmology},
         year = 2019,
        month = may,
       volume = {6},
       number = {1},
          eid = {2},
        pages = {2},
          doi = {10.1186/s40668-019-0028-x},
archivePrefix = {arXiv},
       eprint = {1812.05609},
 primaryClass = {astro-ph.GA},
       adsurl = {https://ui.adsabs.harvard.edu/abs/2019ComAC...6....2N}
}

@ARTICLE{TNGe,
       author = {{Naiman}, Jill P. and {Pillepich}, Annalisa and {Springel}, Volker and
         {Ramirez-Ruiz}, Enrico and {Torrey}, Paul and {Vogelsberger}, Mark and
         {Pakmor}, R{\"u}diger and {Nelson}, Dylan and {Marinacci}, Federico and
         {Hernquist}, Lars and {Weinberger}, Rainer and {Genel}, Shy},
        title = "{First results from the IllustrisTNG simulations: a tale of two elements - chemical evolution of magnesium and europium}",
      journal = {\mnras},
         year = 2018,
        month = jun,
       volume = {477},
       number = {1},
        pages = {1206-1224},
          doi = {10.1093/mnras/sty618},
archivePrefix = {arXiv},
       eprint = {1707.03401},
 primaryClass = {astro-ph.GA},
       adsurl = {https://ui.adsabs.harvard.edu/abs/2018MNRAS.477.1206N}
}

@article{MTNG_02,
       author = {{Hern{\'a}ndez-Aguayo}, C{\'e}sar and {Springel}, Volker and {Pakmor}, R{\"u}diger and {Barrera}, Monica and {Ferlito}, Fulvio and {White}, Simon D.~M. and {Hernquist}, Lars and {Hadzhiyska}, Boryana and {Delgado}, Ana Maria and {Kannan}, Rahul and {Bose}, Sownak and {Frenk}, Carlos},
        title = "{The MillenniumTNG Project: high-precision predictions for matter clustering and halo statistics}",
      journal = {\mnras},
         year = 2023,
        month = sep,
       volume = {524},
       number = {2},
        pages = {2556-2578},
          doi = {10.1093/mnras/stad1657},
archivePrefix = {arXiv},
       eprint = {2210.10059},
 primaryClass = {astro-ph.CO},
       adsurl = {https://ui.adsabs.harvard.edu/abs/2023MNRAS.524.2556H}
}

@article{MTNG_01,
       author = {{Pakmor}, R{\"u}diger and {Springel}, Volker and {Coles}, Jonathan P. and {Guillet}, Thomas and {Pfrommer}, Christoph and {Bose}, Sownak and {Barrera}, Monica and {Delgado}, Ana Maria and {Ferlito}, Fulvio and {Frenk}, Carlos and {Hadzhiyska}, Boryana and {Hern{\'a}ndez-Aguayo}, C{\'e}sar and {Hernquist}, Lars and {Kannan}, Rahul and {White}, Simon D.~M.},
        title = "{The MillenniumTNG Project: the hydrodynamical full physics simulation and a first look at its galaxy clusters}",
      journal = {\mnras},
         year = 2023,
        month = sep,
       volume = {524},
       number = {2},
        pages = {2539-2555},
          doi = {10.1093/mnras/stac3620},
archivePrefix = {arXiv},
       eprint = {2210.10060},
 primaryClass = {astro-ph.CO},
       adsurl = {https://ui.adsabs.harvard.edu/abs/2023MNRAS.524.2539P}
}

@article{MTNG_03,
       author = {{Barrera}, Monica and {Springel}, Volker and {White}, Simon D.~M. and {Hern{\'a}ndez-Aguayo}, C{\'e}sar and {Hernquist}, Lars and {Frenk}, Carlos and {Pakmor}, R{\"u}diger and {Ferlito}, Fulvio and {Hadzhiyska}, Boryana and {Delgado}, Ana Maria and {Kannan}, Rahul and {Bose}, Sownak},
        title = "{The MillenniumTNG Project: semi-analytic galaxy formation models on the past lightcone}",
      journal = {\mnras},
         year = 2023,
        month = nov,
       volume = {525},
       number = {4},
        pages = {6312-6335},
          doi = {10.1093/mnras/stad2688},
archivePrefix = {arXiv},
       eprint = {2210.10419},
 primaryClass = {astro-ph.CO},
       adsurl = {https://ui.adsabs.harvard.edu/abs/2023MNRAS.525.6312B}
}

@article{MTNG_04,
       author = {{Hadzhiyska}, Boryana and {Hernquist}, Lars and {Eisenstein}, Daniel and {Delgado}, Ana Maria and {Bose}, Sownak and {Kannan}, Rahul and {Pakmor}, R{\"u}diger and {Springel}, Volker and {Contreras}, Sergio and {Barrera}, Monica and {Ferlito}, Fulvio and {Hern{\'a}ndez-Aguayo}, C{\'e}sar and {White}, Simon D.~M. and {Frenk}, Carlos},
        title = "{The MillenniumTNG Project: refining the one-halo model of red and blue galaxies at different redshifts}",
      journal = {\mnras},
         year = 2023,
        month = sep,
       volume = {524},
       number = {2},
        pages = {2524-2538},
          doi = {10.1093/mnras/stad279},
archivePrefix = {arXiv},
       eprint = {2210.10068},
 primaryClass = {astro-ph.CO},
       adsurl = {https://ui.adsabs.harvard.edu/abs/2023MNRAS.524.2524H}
}

@article{MTNG_05,
       author = {{Hadzhiyska}, Boryana and {Eisenstein}, Daniel and {Hernquist}, Lars and {Pakmor}, R{\"u}diger and {Bose}, Sownak and {Delgado}, Ana Maria and {Contreras}, Sergio and {Kannan}, Rahul and {White}, Simon D.~M. and {Springel}, Volker and {Frenk}, Carlos and {Hern{\'a}ndez-Aguayo}, C{\'e}sar and {Barrera}, Fulvio Ferlito and {Monica}},
        title = "{The MillenniumTNG Project: an improved two-halo model for the galaxy-halo connection of red and blue galaxies}",
      journal = {\mnras},
         year = 2023,
        month = sep,
       volume = {524},
       number = {2},
        pages = {2507-2523},
          doi = {10.1093/mnras/stad731},
archivePrefix = {arXiv},
       eprint = {2210.10072},
 primaryClass = {astro-ph.CO},
       adsurl = {https://ui.adsabs.harvard.edu/abs/2023MNRAS.524.2507H},
}

@ARTICLE{MTNG_06,
       author = {{Delgado}, Ana Maria and {Hadzhiyska}, Boryana and {Bose}, Sownak and {Springel}, Volker and {Hernquist}, Lars and {Barrera}, Monica and {Pakmor}, R{\"u}diger and {Ferlito}, Fulvio and {Kannan}, Rahul and {Hern{\'a}ndez-Aguayo}, C{\'e}sar and {White}, Simon D.~M. and {Frenk}, Carlos},
        title = "{The MillenniumTNG project: intrinsic alignments of galaxies and haloes}",
      journal = {\mnras},
         year = 2023,
        month = aug,
       volume = {523},
       number = {4},
        pages = {5899-5914},
          doi = {10.1093/mnras/stad1781},
archivePrefix = {arXiv},
       eprint = {2304.12346},
 primaryClass = {astro-ph.CO},
       adsurl = {https://ui.adsabs.harvard.edu/abs/2023MNRAS.523.5899D}
}

@article{MTNG_07,
       author = {{Kannan}, Rahul and {Springel}, Volker and {Hernquist}, Lars and {Pakmor}, R{\"u}diger and {Delgado}, Ana Maria and {Hadzhiyska}, Boryana and {Hern{\'a}ndez-Aguayo}, C{\'e}sar and {Barrera}, Monica and {Ferlito}, Fulvio and {Bose}, Sownak and {White}, Simon D.~M. and {Frenk}, Carlos and {Smith}, Aaron and {Garaldi}, Enrico},
        title = "{The MillenniumTNG project: the galaxy population at z {\ensuremath{\geq}} 8}",
      journal = {\mnras},
         year = 2023,
        month = sep,
       volume = {524},
       number = {2},
        pages = {2594-2605},
          doi = {10.1093/mnras/stac3743},
archivePrefix = {arXiv},
       eprint = {2210.10066},
 primaryClass = {astro-ph.GA},
       adsurl = {https://ui.adsabs.harvard.edu/abs/2023MNRAS.524.2594K}
}

@article{MTNG_08,
       author = {{Bose}, Sownak and {Hadzhiyska}, Boryana and {Barrera}, Monica and {Delgado}, Ana Maria and {Ferlito}, Fulvio and {Frenk}, Carlos and {Hern{\'a}ndez-Aguayo}, C{\'e}sar and {Hernquist}, Lars and {Kannan}, Rahul and {Pakmor}, R{\"u}diger and {Springel}, Volker and {White}, Simon D.~M.},
        title = "{The MillenniumTNG Project: the large-scale clustering of galaxies}",
      journal = {\mnras},
         year = 2023,
        month = sep,
       volume = {524},
       number = {2},
        pages = {2579-2593},
          doi = {10.1093/mnras/stad1097},
archivePrefix = {arXiv},
       eprint = {2210.10065},
 primaryClass = {astro-ph.CO},
       adsurl = {https://ui.adsabs.harvard.edu/abs/2023MNRAS.524.2579B}
}

@ARTICLE{Contreras:2021_SHAMgab,
       author = {{Contreras}, S. and {Angulo}, R.~E. and {Zennaro}, M.},
        title = "{A flexible modelling of galaxy assembly bias}",
      journal = {\mnras},
         year = 2021,
        month = jul,
       volume = {504},
       number = {4},
        pages = {5205-5220},
          doi = {10.1093/mnras/stab1170},
archivePrefix = {arXiv},
       eprint = {2005.03672},
 primaryClass = {astro-ph.GA},
       adsurl = {https://ui.adsabs.harvard.edu/abs/2021MNRAS.504.5205C}
}

@ARTICLE{Contreras:2021_SHAMe,
       author = {{Contreras}, S. and {Angulo}, R.~E. and {Zennaro}, M.},
        title = "{A flexible subhalo abundance matching model for galaxy clustering in redshift space}",
      journal = {\mnras},
         year = 2021,
        month = nov,
       volume = {508},
       number = {1},
        pages = {175-189},
          doi = {10.1093/mnras/stab2560},
archivePrefix = {arXiv},
       eprint = {2012.06596},
 primaryClass = {astro-ph.CO},
       adsurl = {https://ui.adsabs.harvard.edu/abs/2021MNRAS.508..175C}
}

@ARTICLE{Contreras:2019,
       author = {{Contreras}, S. and {Zehavi}, I. and {Padilla}, N. and {Baugh}, C.~M. and
         {Jim{\'e}nez}, E. and {Lacerna}, I.},
        title = "{The evolution of assembly bias}",
      journal = {\mnras},
         year = "2019",
        month = "Mar",
       volume = {484},
       number = {1},
        pages = {1133-1148},
          doi = {10.1093/mnras/stz018},
archivePrefix = {arXiv},
       eprint = {1808.02896},
 primaryClass = {astro-ph.GA},
       adsurl = {https://ui.adsabs.harvard.edu/abs/2019MNRAS.484.1133C}
}

@ARTICLE{Contreras:2023_LIL,
       author = {{Contreras}, Sergio and {Angulo}, Raul E. and {Chaves-Montero}, Jon{\'a}s and {White}, Simon D.~M. and {Aric{\`o}}, Giovanni},
        title = "{Consistent and simultaneous modelling of galaxy clustering and galaxy-galaxy lensing with subhalo abundance matching}",
      journal = {\mnras},
         year = 2023,
        month = mar,
       volume = {520},
       number = {1},
        pages = {489-502},
          doi = {10.1093/mnras/stad122},
archivePrefix = {arXiv},
       eprint = {2211.11745},
 primaryClass = {astro-ph.CO},
       adsurl = {https://ui.adsabs.harvard.edu/abs/2023MNRAS.520..489C}
}

@ARTICLE{Contreras:2023_MTNG,
       author = {{Contreras}, Sergio and {Angulo}, Raul E. and {Springel}, Volker and {White}, Simon D.~M. and {Hadzhiyska}, Boryana and {Hernquist}, Lars and {Pakmor}, R{\"u}diger and {Kannan}, Rahul and {Hern{\'a}ndez-Aguayo}, C{\'e}sar and {Barrera}, Monica and {Ferlito}, Fulvio and {Delgado}, Ana Maria and {Bose}, Sownak and {Frenk}, Carlos},
        title = "{The MillenniumTNG Project: inferring cosmology from galaxy clustering with accelerated N-body scaling and subhalo abundance matching}",
      journal = {\mnras},
         year = 2023,
        month = sep,
       volume = {524},
       number = {2},
        pages = {2489-2506},
          doi = {10.1093/mnras/stac3699},
archivePrefix = {arXiv},
       eprint = {2210.10075},
 primaryClass = {astro-ph.GA},
       adsurl = {https://ui.adsabs.harvard.edu/abs/2023MNRAS.524.2489C}
}

@ARTICLE{Contreras:2023_SDSSlensing,
       author = {{Contreras}, Sergio and {Chaves-Montero}, Jon{\'a}s and {Angulo}, Raul E.},
        title = "{Consistent clustering and lensing of SDSS-III BOSS galaxies with an extended abundance matching formalism}",
      journal = {\mnras},
         year = 2023,
        month = oct,
       volume = {525},
       number = {2},
        pages = {3149-3161},
          doi = {10.1093/mnras/stad2434},
archivePrefix = {arXiv},
       eprint = {2305.09637},
 primaryClass = {astro-ph.CO},
       adsurl = {https://ui.adsabs.harvard.edu/abs/2023MNRAS.525.3149C}
}

@ARTICLE{Contreras:2024,
       author = {{Contreras}, Sergio and {Angulo}, Raul E. and {Chaves-Montero}, Jon{\'a}s and {Kugel}, Roi and {Schaller}, Matthieu and {Schaye}, Joop},
        title = "{Validating the clustering predictions of empirical models with the FLAMINGO simulations}",
      journal = {\aap},
         year = 2024,
        month = oct,
       volume = {690},
          eid = {A311},
        pages = {A311},
          doi = {10.1051/0004-6361/202451671},
archivePrefix = {arXiv},
       eprint = {2407.18912},
 primaryClass = {astro-ph.GA},
       adsurl = {https://ui.adsabs.harvard.edu/abs/2024A&A...690A.311C}
}

@ARTICLE{White:1978,
   author = {{White}, S. D. M. and {Rees}, M. J.},
    title = "{Core condensation in heavy halos - A two-stage theory for galaxy formation and clustering}",
  journal = {\mnras},
archivePrefix = "arXiv",
     year = 1978,
    month = may,
   volume = 183,
    pages = {341-358},
   adsurl = {http://adsabs.harvard.edu/abs/1978MNRAS.183..341W}
}

@ARTICLE{Vale:2006,
   author = {{Vale}, A. and {Ostriker}, J. P.},
    title = "{The non-parametric model for linking galaxy luminosity with halo/subhalo mass}",
  journal = {\mnras},
     year = 2006,
    month = sep,
   volume = 371,
    pages = {1173-1187},
   adsurl = {http://adsabs.harvard.edu/abs/2006MNRAS.371.1173V}
}

@ARTICLE{Conroy:2006,
   author = {{Conroy}, Charlie and {Wechsler}, Risa H. and {Kravtsov}, Andrey V.},
    title = "{Modeling Luminosity-dependent Galaxy Clustering through Cosmic Time}",
  journal = {\apj},
     year = 2006,
    month = aug,
   volume = 647,
    pages = {201-214},
      doi = {10.1086/503602},
   adsurl = {http://adsabs.harvard.edu/abs/2006ApJ...647..201C}
}

@ARTICLE{ChavesMontero:2016,
       author = {{Chaves-Montero}, Jon{\'a}s and {Angulo}, Raul E. and {Schaye}, Joop and {Schaller}, Matthieu and {Crain}, Robert A. and {Furlong}, Michelle and {Theuns}, Tom},
        title = "{Subhalo abundance matching and assembly bias in the EAGLE simulation}",
      journal = {\mnras},
         year = 2016,
        month = aug,
       volume = {460},
       number = {3},
        pages = {3100-3118},
          doi = {10.1093/mnras/stw1225},
archivePrefix = {arXiv},
       eprint = {1507.01948},
 primaryClass = {astro-ph.GA},
       adsurl = {https://ui.adsabs.harvard.edu/abs/2016MNRAS.460.3100C}
}

@ARTICLE{Ortega:2024,
       author = {{Ortega-Martinez}, S. and {Contreras}, S. and {Angulo}, R.},
        title = "{SHAMe-SF: Predicting the clustering of star-forming galaxies with an enhanced abundance matching model}",
      journal = {\aap},
         year = 2024,
        month = sep,
       volume = {689},
          eid = {A66},
        pages = {A66},
          doi = {10.1051/0004-6361/202449597},
archivePrefix = {arXiv},
       eprint = {2401.17374},
 primaryClass = {astro-ph.CO},
       adsurl = {https://ui.adsabs.harvard.edu/abs/2024A&A...689A..66O}
}

@ARTICLE{Rocher:2023,
       author = {{Rocher}, Antoine and {Ruhlmann-Kleider}, Vanina and {Burtin}, Etienne and {Yuan}, Sihan and {de Mattia}, Arnaud and {Ross}, Ashley J. and {Aguilar}, Jessica and {Ahlen}, Steven and {Alam}, Shadab and {Bianchi}, Davide and {Brooks}, David and {Cole}, Shaun and {Dawson}, Kyle and {de la Macorra}, Axel and {Doel}, Peter and {Eisenstein}, Daniel J. and {Fanning}, Kevin and {Forero-Romero}, Jaime E. and {Garrison}, Lehman H. and {Gontcho A Gontcho}, Satya and {Gonzalez-Perez}, Violeta and {Guy}, Julien and {Hadzhiyska}, Boryana and {Hahn}, ChangHoon and {Honscheid}, Klaus and {Kisner}, Theodore and {Landriau}, Martin and {Lasker}, James and {E. Levi}, Michael and {Manera}, Marc and {Meisner}, Aaron and {Miquel}, Ramon and {Moustakas}, John and {Mueller}, Eva-Maria and {Newman}, Jeffrey A. and {Nie}, Jundan and {Percival}, Will J. and {Poppett}, Claire and {Qin}, Fei and {Rossi}, Graziano and {Samushia}, Lado and {Sanchez}, Eusebio and {Schlegel}, David and {Schubnell}, Michael and {Seo}, Hee-Jong and {Tarl{\'e}}, Gregory and {Vargas-Maga{\~n}a}, Mariana and {Weaver}, Benjamin A. and {Yu}, Jiaxi and {Zhang}, Hanyu and {Zheng}, Zheng and {Zhou}, Zhimin and {Zou}, Hu},
        title = "{The DESI One-Percent survey: exploring the Halo Occupation Distribution of Emission Line Galaxies with ABACUSSUMMIT simulations}",
      journal = {\jcap},
         year = 2023,
        month = oct,
       volume = {2023},
       number = {10},
          eid = {016},
        pages = {016},
          doi = {10.1088/1475-7516/2023/10/016},
archivePrefix = {arXiv},
       eprint = {2306.06319},
 primaryClass = {astro-ph.CO},
       adsurl = {https://ui.adsabs.harvard.edu/abs/2023JCAP...10..016R}
}

@ARTICLE{Ortega:2025,
       author = {{Ortega-Martinez}, Sara and {Contreras}, Sergio and {Angulo}, Raul E. and {Chaves-Montero}, Jon{\'a}s},
        title = "{Investigating the galaxy{\textendash}halo connection of DESI emission-line galaxies with SHAMe-SF}",
      journal = {\aap},
         year = 2025,
        month = may,
       volume = {697},
          eid = {A226},
        pages = {A226},
          doi = {10.1051/0004-6361/202453086},
archivePrefix = {arXiv},
       eprint = {2411.11830},
 primaryClass = {astro-ph.CO},
       adsurl = {https://ui.adsabs.harvard.edu/abs/2025A&A...697A.226O}
}

@ARTICLE{Mahony:2025,
       author = {{Mahony}, Constance and {Contreras}, Sergio and {Angulo}, Raul E. and {Alonso}, David and {Georgiou}, Christos and {Dvornik}, Andrej},
        title = "{Cosmological constraints from galaxy clustering and galaxy-galaxy lensing with extended SubHalo Abundance Matching}",
      journal = {arXiv e-prints},
         year = 2025,
        month = jul,
          eid = {arXiv:2507.01601},
        pages = {arXiv:2507.01601},
          doi = {10.48550/arXiv.2507.01601},
archivePrefix = {arXiv},
       eprint = {2507.01601},
 primaryClass = {astro-ph.CO},
       adsurl = {https://ui.adsabs.harvard.edu/abs/2025arXiv250701601M}
}

@ARTICLE{EuclidFlagship01,
       author = {{Euclid Collaboration} and {Castander}, F.~J. and {Fosalba}, P. and {Stadel}, J. and {Potter}, D. and {Carretero}, J. and {Tallada-Cresp{\'\i}}, P. and {Pozzetti}, L. and {Bolzonella}, M. and {Mamon}, G.~A. and {Blot}, L. and {Hoffmann}, K. and {Huertas-Company}, M. and {Monaco}, P. and {Gonzalez}, E.~J. and {De Lucia}, G. and {Scarlata}, C. and {Breton}, M. -A. and {Linke}, L. and {Viglione}, C. and {Li}, S. -S. and {Zhai}, Z. and {Baghkhani}, Z. and {Pardede}, K. and {Neissner}, C. and {Teyssier}, R. and {Crocce}, M. and {Tutusaus}, I. and {Miller}, L. and {Congedo}, G. and {Biviano}, A. and {Hirschmann}, M. and {Pezzotta}, A. and {Aussel}, H. and {Hoekstra}, H. and {Kitching}, T. and {Percival}, W.~J. and {Guzzo}, L. and {Mellier}, Y. and {Oesch}, P.~A. and {Bowler}, R.~A.~A. and {Bruton}, S. and {Allevato}, V. and {Gonzalez-Perez}, V. and {Manera}, M. and {Avila}, S. and {Kov{\'a}cs}, A. and {Aghanim}, N. and {Altieri}, B. and {Amara}, A. and {Amendola}, L. and {Andreon}, S. and {Auricchio}, N. and {Baccigalupi}, C. and {Baldi}, M. and {Balestra}, A. and {Bardelli}, S. and {Bender}, R. and {Bernardeau}, F. and {Bodendorf}, C. and {Bonino}, D. and {Branchini}, E. and {Brescia}, M. and {Brinchmann}, J. and {Camera}, S. and {Capobianco}, V. and {Carbone}, C. and {Casas}, S. and {Castellano}, M. and {Castignani}, G. and {Cavuoti}, S. and {Cimatti}, A. and {Colodro-Conde}, C. and {Conselice}, C.~J. and {Conversi}, L. and {Copin}, Y. and {Corcione}, L. and {Courbin}, F. and {Courtois}, H.~M. and {Da Silva}, A. and {Degaudenzi}, H. and {Di Giorgio}, A.~M. and {Dinis}, J. and {Douspis}, M. and {Dubath}, F. and {Duncan}, C.~A.~J. and {Dupac}, X. and {Dusini}, S. and {Ealet}, A. and {Farina}, M. and {Farrens}, S. and {Ferriol}, S. and {Fotopoulou}, S. and {Fourmanoit}, N. and {Frailis}, M. and {Franceschi}, E. and {Franzetti}, P. and {Galeotta}, S. and {Gillard}, W. and {Gillis}, B. and {Giocoli}, C. and {G{\'o}mez-Alvarez}, P. and {Granett}, B.~R. and {Grazian}, A. and {Grupp}, F. and {Haugan}, S.~V.~H. and {Holliman}, M.~S. and {Holmes}, W. and {Hook}, I. and {Hormuth}, F. and {Hornstrup}, A. and {Hudelot}, P. and {Ili{\'c}}, S. and {Jahnke}, K. and {Jhabvala}, M. and {Joachimi}, B. and {Keih{\"a}nen}, E. and {Kermiche}, S. and {Kiessling}, A. and {Kilbinger}, M. and {Kohley}, R. and {Kubik}, B. and {K{\"u}mmel}, M. and {Kunz}, M. and {Kurki-Suonio}, H. and {Lahav}, O. and {Laureijs}, R. and {Le Mignant}, D. and {Liebing}, P. and {Ligori}, S. and {Lilje}, P.~B. and {Lindholm}, V. and {Lloro}, I. and {Maino}, D. and {Maiorano}, E. and {Mansutti}, O. and {Marcin}, S. and {Marggraf}, O. and {Markovic}, K. and {Martinelli}, M. and {Martinet}, N. and {Marulli}, F. and {Massey}, R. and {Masters}, D.~C. and {Maurogordato}, S. and {McCracken}, H.~J. and {Medinaceli}, E. and {Mei}, S. and {Melchior}, M. and {Meneghetti}, M. and {Merlin}, E. and {Meylan}, G. and {Mohr}, J.~J. and {Moresco}, M. and {Moscardini}, L. and {Munari}, E. and {Nakajima}, R. and {Nichol}, R.~C. and {Niemi}, S. -M. and {Padilla}, C. and {Paech}, K. and {Paltani}, S. and {Pasian}, F. and {Peacock}, J.~A. and {Pedersen}, K. and {Pettorino}, V. and {Pires}, S. and {Polenta}, G. and {Poncet}, M. and {Popa}, L.~A. and {Raison}, F. and {Rebolo}, R. and {Renzi}, A. and {Rhodes}, J. and {Riccio}, G. and {Romelli}, E. and {Roncarelli}, M. and {Rosset}, C. and {Rossetti}, E. and {Rusholme}, B. and {Saglia}, R. and {Sakr}, Z. and {S{\'a}nchez}, A.~G. and {Sapone}, D. and {Schewtschenko}, J.~A. and {Schirmer}, M. and {Schneider}, P. and {Schrabback}, T. and {Scodeggio}, M. and {Secroun}, A. and {Sefusatti}, E. and {Seidel}, G. and {Serrano}, S. and {Sirignano}, C. and {Sirri}, G. and {Stanco}, L. and {Starck}, J. -L. and {Steinwagner}, J. and {Taylor}, A.~N. and {Teplitz}, H.~I.},
        title = "{Euclid: V. The Flagship galaxy mock catalogue: A comprehensive simulation for the Euclid mission}",
      journal = {\aap},
         year = 2025,
        month = may,
       volume = {697},
          eid = {A5},
        pages = {A5},
          doi = {10.1051/0004-6361/202450853},
archivePrefix = {arXiv},
       eprint = {2405.13495},
 primaryClass = {astro-ph.CO},
       adsurl = {https://ui.adsabs.harvard.edu/abs/2025A&A...697A...5E}
}

@ARTICLE{ABACUSSUMMIT,
       author = {{Maksimova}, Nina A. and {Garrison}, Lehman H. and {Eisenstein}, Daniel J. and {Hadzhiyska}, Boryana and {Bose}, Sownak and {Satterthwaite}, Thomas P.},
        title = "{ABACUSSUMMIT: a massive set of high-accuracy, high-resolution N-body simulations}",
      journal = {\mnras},
         year = 2021,
        month = dec,
       volume = {508},
       number = {3},
        pages = {4017-4037},
          doi = {10.1093/mnras/stab2484},
archivePrefix = {arXiv},
       eprint = {2110.11398},
 primaryClass = {astro-ph.CO},
       adsurl = {https://ui.adsabs.harvard.edu/abs/2021MNRAS.508.4017M}
}

@ARTICLE{Orsi:2018,
       author = {{Orsi}, {\'A}lvaro A. and {Angulo}, Ra{\'u}l E.},
        title = "{The impact of galaxy formation on satellite kinematics and redshift-space distortions}",
      journal = {\mnras},
         year = 2018,
        month = apr,
       volume = {475},
       number = {2},
        pages = {2530-2544},
          doi = {10.1093/mnras/stx3349},
archivePrefix = {arXiv},
       eprint = {1708.00956},
 primaryClass = {astro-ph.CO},
       adsurl = {https://ui.adsabs.harvard.edu/abs/2018MNRAS.475.2530O}
}

@article{Gadget-4,
    author = {{Springel}, Volker and {Pakmor}, R{\"u}diger and {Zier}, Oliver and {Reinecke}, Martin},
    title = "{Simulating cosmic structure formation with the GADGET-4 code}",
    journal = "MNRAS",
    year = 2021,
    month = sep,
    volume = {506},
    number = {2},
    pages = {2871-2949},
    doi = {10.1093/mnras/stab1855},
    archivePrefix = {arXiv},
    eprint = {2010.03567},
    primaryClass = {astro-ph.IM}
}

@ARTICLE{Angulo:2016,
       author = {{Angulo}, Raul E. and {Pontzen}, Andrew},
        title = "{Cosmological N-body simulations with suppressed variance}",
      journal = {\mnras},
         year = "2016",
        month = "Oct",
       volume = {462},
       number = {1},
        pages = {L1-L5},
          doi = {10.1093/mnrasl/slw098},
archivePrefix = {arXiv},
       eprint = {1603.05253},
 primaryClass = {astro-ph.CO},
       adsurl = {https://ui.adsabs.harvard.edu/abs/2016MNRAS.462L...1A}
}

@ARTICLE{scipy,
  author  = {Virtanen, Pauli and Gommers, Ralf and Oliphant, Travis E. and
            Haberland, Matt and Reddy, Tyler and Cournapeau, David and
            Burovski, Evgeni and Peterson, Pearu and Weckesser, Warren and
            Bright, Jonathan and {van der Walt}, St{\'e}fan J. and
            Brett, Matthew and Wilson, Joshua and Millman, K. Jarrod and
            Mayorov, Nikolay and Nelson, Andrew R. J. and Jones, Eric and
            Kern, Robert and Larson, Eric and Carey, C J and
            Polat, {\.I}lhan and Feng, Yu and Moore, Eric W. and
            {VanderPlas}, Jake and Laxalde, Denis and Perktold, Josef and
            Cimrman, Robert and Henriksen, Ian and Quintero, E. A. and
            Harris, Charles R. and Archibald, Anne M. and
            Ribeiro, Ant{\^o}nio H. and Pedregosa, Fabian and
            {van Mulbregt}, Paul and {SciPy 1.0 Contributors}},
  title   = {{{SciPy} 1.0: Fundamental Algorithms for Scientific
            Computing in Python}},
  journal = {Nature Methods},
  year    = {2020},
  volume  = {17},
  pages   = {261--272},
  adsurl  = {https://rdcu.be/b08Wh},
  doi     = {10.1038/s41592-019-0686-2},
}

\begin{appendix}

\section{Mimicking baryonic effects in a dark-matter-only simulation}
\label{sec:Mimicking}

\begin{figure*}
\includegraphics[width=0.95\textwidth]{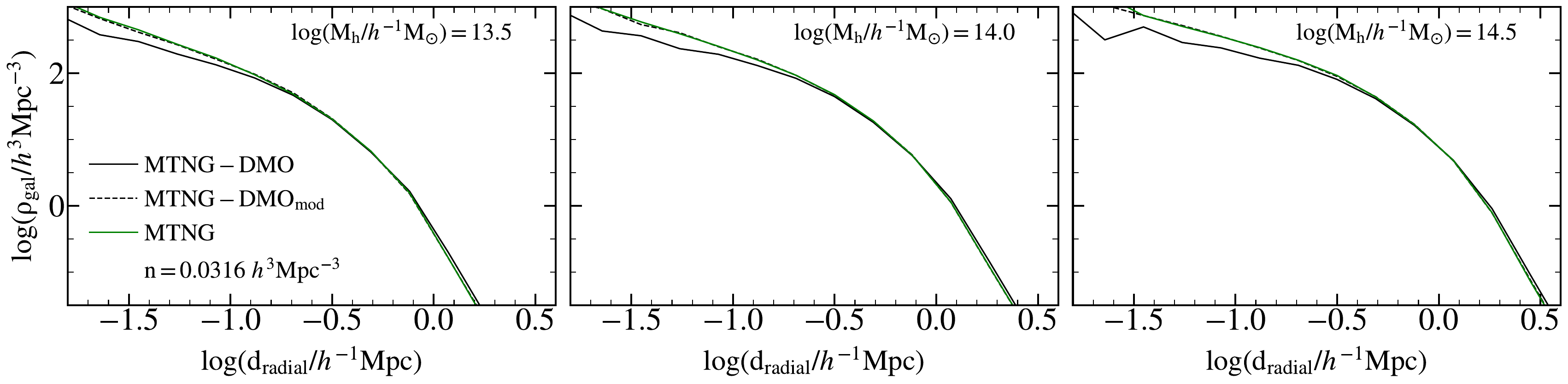}
\includegraphics[width=0.95\textwidth]{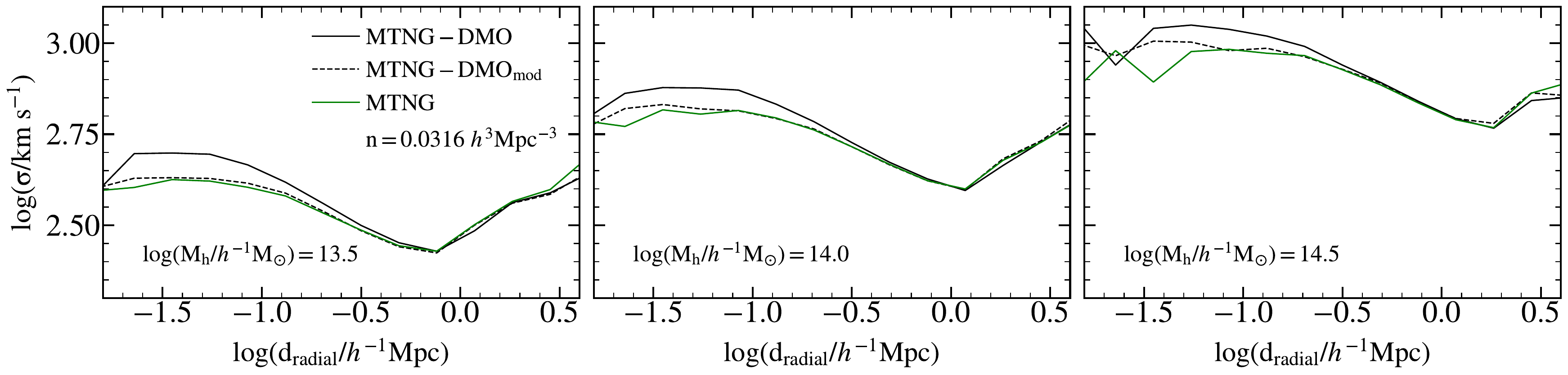}
\caption{Similar to Figs.~\ref{Fig:prof_dist} and \ref{Fig:prof_vel}, but including the predictions of the \MTNGdmoMOD\ simulation (dashed lines), in which the positions and velocities of subhaloes were modified to mimic those of the hydrodynamic \MTNG\ simulation.}
\label{Fig:fit}
\end{figure*}

In this appendix, we detail the algorithm used to modify the positions and velocities of subhaloes in \MTNGdmo\ so as to reproduce the baryonic effects present in the hydrodynamic \MTNG\ run. The procedure is divided into three steps. We first modify the positions of satellite subhaloes, then adjust their velocities, and finally correct the velocities of central galaxies.

\subsection*{Modifying the satellite positions}

To modify satellite positions, we first compute the cumulative number of satellites as a function of their distance to the halo centre, normalised by the virial radius $r_{200c}$. We normalise this distribution by the total number of satellites per halo, which restricts the function to the interval $[0,1]$:
\begin{equation}
    {\rm \phi_c}(d_{\rm radial}/r_{200c}) = \frac{N_{\rm sat}(< d_{\rm radial}/r_{200c})}{N_{\rm sat,tot}},
\end{equation}
where $N_{\rm sat}(< d_{\rm radial}/r_{200c})$ is the number of satellites within $d_{\rm radial}$ and $N_{\rm sat,tot}$ is the total number of satellites in the halo.

Normalising distances by $r_{200c}$ and the counts by $N_{\rm sat,tot}$ yields a distribution with very weak halo-mass dependence. We then compute the difference between the inverse cumulative functions of the hydrodynamic and dark-matter-only simulations, i.e. the difference in the average satellite distance at fixed cumulative satellite fraction:
\begin{equation}
    \Delta r_{\rm sat}(n_{\rm sat,rad}) = \phi^{I}_{c,{\rm hydro}}(n_{\rm sat,rad}) - \phi^{I}_{c,{\rm dmo}}(n_{\rm sat,rad}),
\end{equation}
where $\Delta r_{\rm sat}(n_{\rm sat,rad})$ is expressed in units of the virial radius and
\begin{equation}
    n_{\rm sat,rad} = \frac{N_{\rm sat}(< d_{\rm radial}/r_{200c})}{N_{\rm sat,tot}}
\end{equation}
is the satellite rank within the halo (i.e. $n_{\rm sat,rad}=1$ for the outermost satellite and $n_{\rm sat,rad}=1/N_{\rm sat,tot}$ for the innermost satellite).

We find that $\Delta r_{\rm sat}(n_{\rm sat,rad})$ shows almost no dependence on host-halo mass. We therefore compute an average $\Delta r_{\rm sat}(n_{\rm sat,rad})$ combining all haloes with $M_{\rm h} > 10^{13}\,\hMsun$. This is done using only the matched subhaloes (about 97\% of the total sample for our highest number-density case). Finally, we modify the satellite distances as
\begin{equation}
    d'_{\rm radial} = d_{\rm radial} + \Delta r_{\rm sat}(n_{\rm sat,rad}).
\end{equation}
The correction is applied only to the modulus of the distance to the halo centre; we do not change the direction of the position vector, nor do we alter the spatial alignment of satellites.

We calibrated $\Delta r_{\rm sat}(n_{\rm sat,rad})$ using the sample with number density $0.0316~\ihMpcC$, which shows the lowest noise. The resulting satellite distributions for \MTNGdmoMOD, \MTNG, and \MTNGdmo\ are shown in the upper panel of Fig.~\ref{Fig:fit}. This correction reproduces well the radial distributions over the full halo-mass range. We also checked the median satellite distance as a function of halo mass (analogous to Fig.~\ref{Fig:mean_dist}) and found a similar level of accuracy (not shown).

\subsection*{Modifying the satellite velocities}

We now turn to satellite velocities. We start by measuring, in both simulations, the mean modulus of the satellite velocity relative to the host halo, as a function of the satellite distance to the halo centre and in bins of halo mass. At this point, we use the satellite positions already corrected as described above. It is essential to correct positions first, because satellite velocity strongly correlates with radius; if the order is inverted, the median velocity of the final sample will not match that of the hydrodynamic run.

We denote the average satellite speed in a given simulation as
\[
    |v|(d_{\rm radial}, M_{\rm h}) \equiv \langle | \mathbf{v}_{\rm sat} - \mathbf{v}_{\rm h} | \rangle,
\]
and consider the ratio between the hydrodynamic and dark-matter-only simulations. We find that, at small radii, this ratio is $\sim 0.9$, while at large radii it approaches 1, with a smooth transition between both regimes. To model this behaviour and apply it to all subhaloes, we describe the ratio with an error-function-like parametrisation:
\begin{equation}
\begin{aligned}
   \frac{|v_{\rm hydro}|}{|v_{\rm dmo}|}(d_{\rm radial}) = &\ \frac{(f_{\max} - f_{\min})}{2}\,
   {\rm erf}\left(\frac{\log(d_{\rm radial}) - \log(d_0)}{\sigma}\right) \\
   &+ \frac{(f_{\max} + f_{\min})}{2},
\end{aligned}
\end{equation}
where $f_{\min}$ and $f_{\max}$ are the minimum and maximum values of the ratio, $d_0$ is the characteristic radius of the transition, $\sigma$ controls its smoothness, and ${\rm erf}$ is the error function,
\begin{equation}
    {\rm erf}(x) = \frac{2}{\sqrt{\pi}} \int_0^x e^{-t^2} \, {\rm d}t.
\end{equation}

We find that $f_{\min}$ and $f_{\max}$ show a weak dependence on halo mass, which we model as
\begin{equation}
    f_{\min}(M_{\rm h}) = a_{f,\min}\,(\log M_{\rm h} - 14) + f_{\min,14},
\end{equation}
and
\begin{equation}
    f_{\max}(M_{\rm h}) = a_{f,\max}\,(\log M_{\rm h} - 14) + f_{\max,14},
\end{equation}
where $f_{\min,14}$ and $f_{\max,14}$ are the values for haloes of $10^{14}\,\hMsun$, and $a_{f,\min}$ and $a_{f,\max}$ encode the mild mass dependence.

For the \MTNG\ hydrodynamic simulation we obtain the following best-fitting parameters:
\[
    (f_{\min,14},\, f_{\max,14},\, \sigma,\, \log d_0,\, a_{f,\min},\, a_{f,\max})
    =
\]
\[
    (0.8733,\ 1.004,\ 0.285,\ -0.715,\ 0.05,\ 0.03).
\]
The calibration was done using the highest number-density subhalo sample, which also agrees well with the lower-density samples.

The lower panel of Fig.~\ref{Fig:fit} shows the velocity-dispersion profiles for three halo-mass bins, comparing \MTNG, \MTNGdmo, and \MTNGdmoMOD\ (grey dashed line). The corrected catalogue reproduces the hydrodynamic velocity profiles very well. We also tested a simplified correction that ignores the halo-mass dependence; this has negligible impact on both the velocity profiles and the clustering results of Fig.~\ref{Fig:clustering}.

\subsection*{Change in the velocities of centrals}

\begin{figure}
\includegraphics[width=0.45\textwidth]{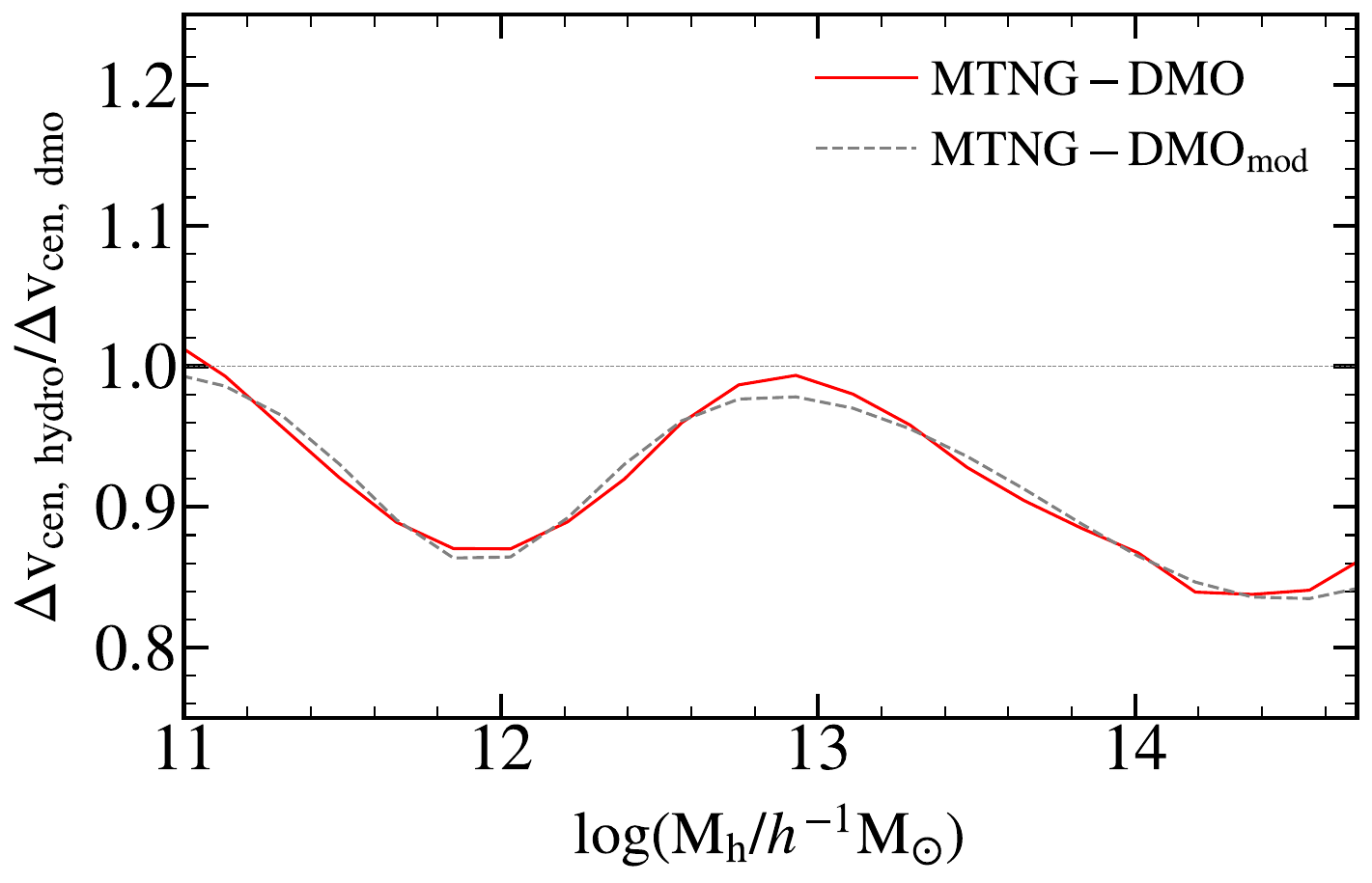}
\caption{Ratio of the relative velocities of central subhaloes with respect to their host haloes between the \MTNG\ and \MTNGdmo\ simulations, as a function of host-halo mass. The grey dashed line shows a two-Gaussian parametrisation used to describe this relation and to correct dark-matter-only simulations.}
\label{Fig:VelBias}
\end{figure}

\begin{figure*}
\includegraphics[width=0.95\textwidth]{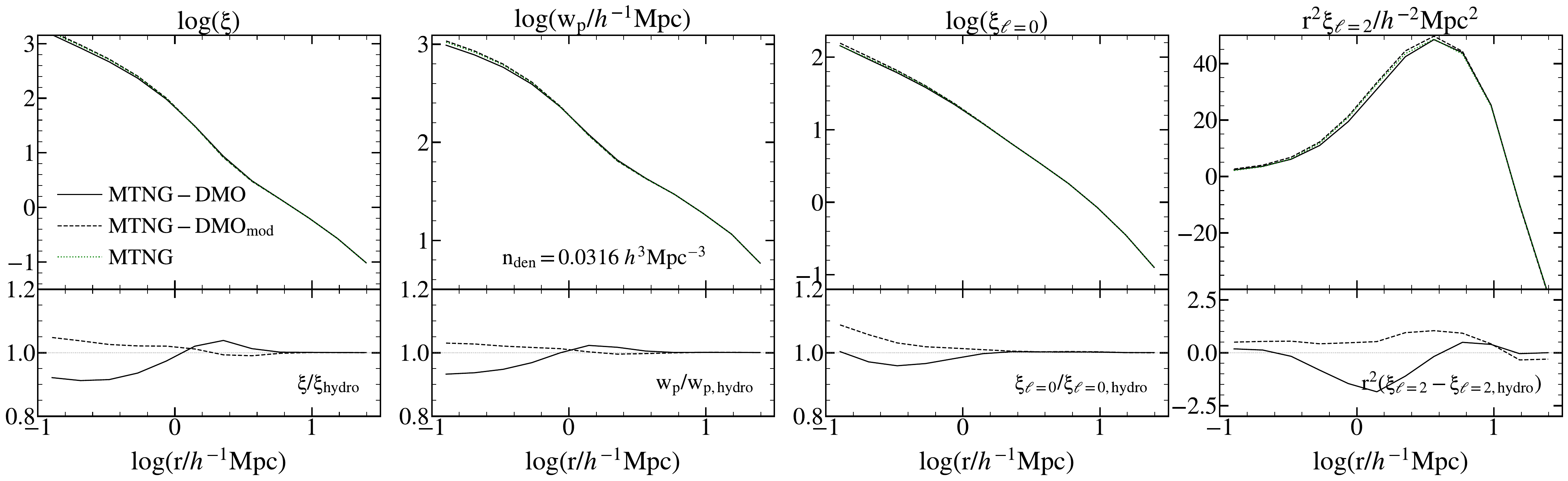}
\caption{Clustering measurements for the \MTNG, \MTNGdmo, and the modified \MTNGdmo\ version (labelled \MTNGdmoMOD). All results use the matched catalogues selected by a $\vpeak$ threshold corresponding to a number density of $0.0316\ \ihMpcC$.}
\label{Fig:ClusterTest}
\end{figure*}

The subhalo finder \texttt{SUBFIND} assigns the same position to the central subhalo/galaxy and to the host halo (the potential minimum). Velocities, however, are usually defined as the mass-weighted mean of all particles or cells in the halo, so the velocity of the central object differs from that of the halo. This offset, known as velocity bias, scales with the halo velocity dispersion and can noticeably affect subhalo/galaxy clustering \citep{Guo:2015}. Although a detailed study of baryonic effects on centrals is beyond the scope of this paper, we must correct for this velocity difference to isolate the impact of baryons on satellite clustering.

To do so, we extend our halo matching down to $10^{11}\,\hMsun$, the minimum halo mass of our densest sample. Matching centrals between simulations with identical initial conditions is straightforward. We then compute, for each matched halo, the relative velocity of the central object with respect to the halo (central minus halo velocity) in both simulations and take the ratio. Figure~\ref{Fig:VelBias} shows this ratio as a function of halo mass. The largest differences occur around $10^{12}\,\hMsun$ and $10^{14.5}\,\hMsun$, while intermediate masses show smaller offsets.

We parametrise this behaviour as
\begin{equation}
  g(M_{\rm h}) = 1 - G_1(\log M_{\rm h}) - G_2(\log M_{\rm h}),
\end{equation}
where $G_1$ and $G_2$ are Gaussian functions,
\begin{equation}
    G(x; a, b, c) = a\, \exp\left[-\frac{(x-b)^2}{2c^2}\right].
\end{equation}
The best-fitting parameters are
\[
(a_1, b_1, c_1) = (0.139, 11.93, 0.376), \quad
\]
\[
(a_2, b_2, c_2) = (0.166, 14.48, 0.732).
\]
As shown in Fig.~\ref{Fig:VelBias}, this parametrisation reproduces the velocity ratios over the full halo-mass range. We apply $g(M_{\rm h})$ as a multiplicative correction to the central velocities of all haloes in the dark-matter-only simulation.

All the corrections above should be regarded as first-order. We did not explore possible dependencies on secondary halo properties (concentration, formation time, or environment), nor did we test whether the functional form persists for other galaxy-formation prescriptions. Nonetheless, the methodology can be used as a basis for a more accurate and universal implementation of baryonic effects in subhaloes.

As a proof-of-concept, we apply all corrections to the matched catalogue of the \MTNGdmo\ simulation and compare its clustering to that of the hydrodynamic \MTNG\ run (Fig.~\ref{Fig:ClusterTest}). We find very good agreement between the modified \MTNGdmo\ and the full \MTNG\ simulation, with only minor differences at small scales in the multipoles of the correlation function. Although developing a full baryonification model for subhaloes is beyond the scope of this work, our results demonstrate that the relations presented here provide a solid foundation for constructing a more general and physically motivated subhalo baryonification framework.

\section{Dependence on the galaxy-formation physics}
\label{sec:zoom}

\begin{figure}
\includegraphics[width=0.45\textwidth]{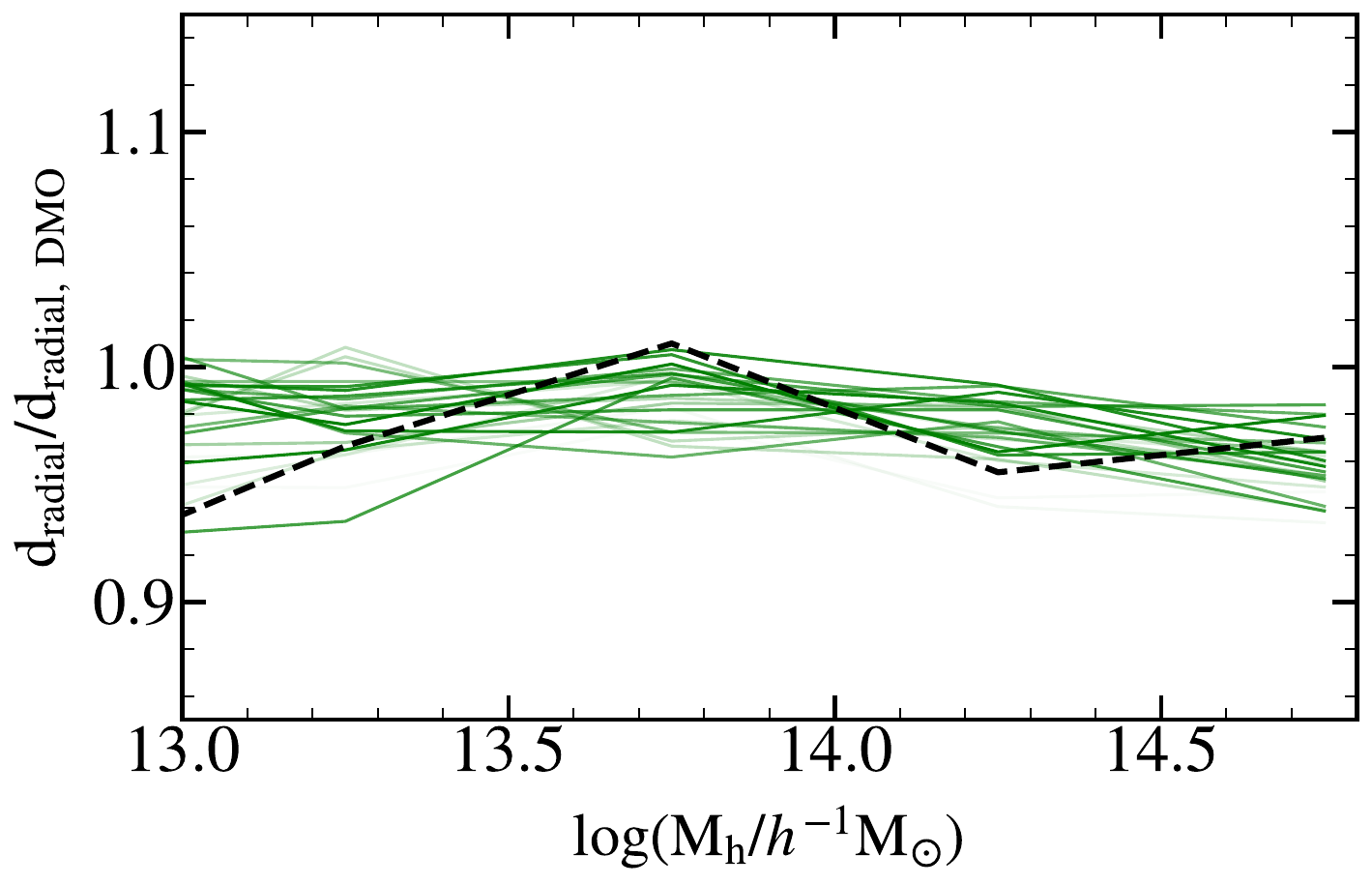}
\includegraphics[width=0.45\textwidth]{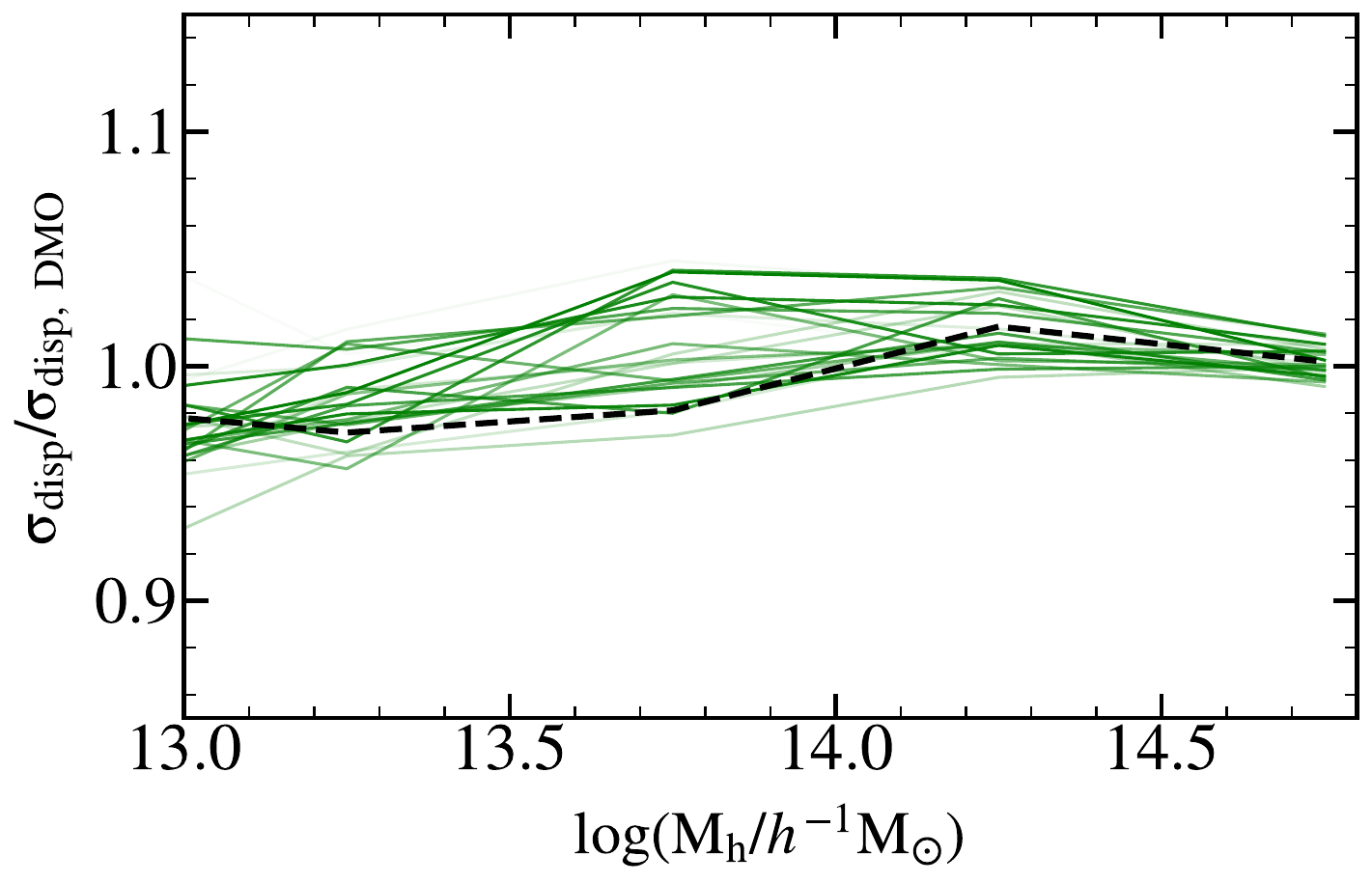}
\caption{Top: ratio of the median satellite distance to the halo centre as a function of halo mass between a suite of zoom-in hydrodynamic simulations and their dark-matter-only counterparts. The dark-matter-only haloes are selected with a $\vpeak$ cut equivalent to a number density of ${\rm n_{den} = 0.1~} \ihMpcC$ in the full \MTNGdmo\ box. The zoom-in with the same physical model as \MTNG\ is shown as a black dashed line. Green lines show runs with varied galaxy-formation parameters; lighter (darker) colours correspond to lower (higher) values of {\tt WindEnergyIn1e51erg}, the parameter that correlates best with the amplitude of the effect. Bottom: same as the top panel, but for the satellite velocity dispersion.}
\label{Fig:zoom}
\end{figure}

\begin{table*}
\caption{Astrophysical parameters varied in the zoom-in simulations, along with the lower and upper limits and the fiducial value. The parameters are: {\tt MaxSfrTimescale}, the star-formation timescale at the threshold density; {\tt WindEnergyIn1e51erg}, the normalisation of the energy in galactic winds per unit star formation; {\tt VariableWindVelFactor}, the normalisation of the galactic wind speed; {\tt WindFreeTravelDensFac}, the gas-density factor that controls when collisionless wind particles recouple to the hydrodynamics; {\tt BlackHoleFeedbackFactor}, the normalisation of AGN feedback energy in the high-accretion state; {\tt RadioFeedbackFactor}, the normalisation of AGN feedback in the low-accretion state; and {\tt RadioFeedbackReorientationFactor}, the normalisation of the frequency of AGN feedback events in the low-accretion state.}
\begin{tabular}{ |c|c|c|c| }
 \hline
 Parameter & min. value & fid. value & max. value\\
 \hline
 MaxSfrTimescale   & $1.19\times10^{-3}$  & $2.27\times10^{-3}$ & $4.48\times10^{-3}$\\
 WindEnergyIn1e51erg & $1.25$ & $3.6$ & $1.44\times10^{1}$\\
 VariableWindVelFactor & $3.73$ & $7.4$ & $1.46\times10^{1}$\\
 WindFreeTravelDensFac & $5.81\times10^{-3}$ & $5.23\times10^{-2}$ & $5\times10^{-1}$\\
 BlackHoleFeedbackFactor & $5.23\times10^{-2}$ & $1\times10^{-1}$ & $1.96\times10^{-1}$\\
 RadioFeedbackFactor & $1.29\times10^{-3}$ & $1$ & $1.46$\\
 RadioFeedbackReiorientationFactor & $1.06\times10^{1}$ & $2\times10^{1}$ & $3.95\times10^{1}$\\
 \hline
\end{tabular}
\label{Table:ZoomParams}
\end{table*}

In this paper, we used the \MTNG\ hydrodynamic simulation to quantify the effects of baryons on satellite positions and velocities and on galaxy clustering. While the trends we found may qualitatively hold for other galaxy-formation models, the amplitude of baryonic effects in the real Universe is not well constrained, and it is not obvious how baryonic effects on the matter distribution translate into baryonic effects on satellites. To explore the possible range of amplitudes, we analysed a suite of 26 zoom-in simulations that share the same initial conditions as \MTNG\ but adopt variation of the galaxy-formation prescriptions. The initial conditions were produced using a multi-zoom technique described in \cite{Burger:2025}. Each of these simulations re-simulates 452 haloes at the same resolution as \MTNG.

The zoom-in simulations were run with \texttt{AREPO} and adopt the same numerical setup, but differ in seven astrophysical parameters varied over broad ranges (Table~\ref{Table:ZoomParams}). These parameters span feedback strengths, wind energetics, and star-formation efficiencies, and were chosen to cover a plausible range of galaxy-formation physics. This ensemble represents the first stage of a larger zoom-in programme (Maion et al., in prep.) and can be used to estimate an upper limit on baryonic effects in satellites.

We match haloes and subhaloes in each zoom-in run to those in the corresponding dark-matter-only zoom-in simulation following the same procedure as in Section~\ref{sec:matched}. Each simulation contains 29 haloes above $10^{13}\,\hMsun$, which limits the statistical power of the analysis. To increase the signal, we consider all subhaloes with $\vpeak \gtrsim 76~{\rm km\,s^{-1}}$, corresponding to a number density of $0.1~\ihMpcC$ in the full \MTNGdmo\ box.

Figure~\ref{Fig:zoom} shows the ratio of the median satellite distance (top) and the ratio of the satellite velocity dispersion (bottom) between the zoom-in hydrodynamic runs and their dark-matter-only counterparts, analogous to Figs.~\ref{Fig:mean_dist} and \ref{Fig:mean_vel}. The zoom-in run with the fiducial \MTNG\ physics is shown in black. We do not find strong trends with most galaxy-formation parameters, although there are weak correlations with parameters linked to galactic winds and to the star-formation timescale. In Fig.~\ref{Fig:zoom}, lighter (darker) green lines indicate runs with lower (higher) values of {\tt WindEnergyIn1e51erg}. There is a mild tendency for lower wind-energy models to place satellites closer to halo centres and to reduce their velocities consistent with expectations from enhanced orbit decay with increased baryonic density in the CGM, but the scatter is too large to draw firm conclusions.

The small number of massive haloes also limits our ability to constrain the maximum possible impact of baryons on satellites. We do see that baryonic effects can be up to a factor of $\sim 2$ larger than in the fiducial model. Although part of this may be due to noise, this factor provides a reasonable first-order upper bound on how much baryons could shift satellite positions or velocities in other galaxy-formation scenarios. Because of the limited sample, we did not attempt to measure more demanding statistics. In future work, we will explore these dependencies using larger suites of hydrodynamic simulations.

\end{appendix}

\end{document}